\documentclass[twocolumn,showpacs,preprintnumbers,amsmath,amssymb]{revtex4}

\usepackage{pifont}

\usepackage{graphicx}
\usepackage{dcolumn}
\usepackage{bm}
\usepackage[colorlinks,linkcolor=blue,anchorcolor=blue,citecolor=blue]{hyperref}
\usepackage{lineno} 
\usepackage{amssymb}
\usepackage{amsmath}
\usepackage{latexsym}
\usepackage{mathrsfs}

\begin{document}
\preprint{}

\title{Gaussian Process and L\'evy Walk under Stochastic Non-instantaneous Resetting and Stochastic Rest}

\author{Tian Zhou$^{1}$}
\author{Pengbo Xu$^{2}$}
\author{Weihua Deng$^{1}$}
\affiliation{$^1$School of Mathematics and Statistics, Gansu Key Laboratory of Applied Mathematics and Complex Systems, Lanzhou University, Lanzhou 730000, P.R. China}
\affiliation{$^2$School of Mathematical Sciences, Peking University, Beijing 100871, P.R. China}


\begin{abstract}
A stochastic process with movement, return, and rest phases is considered in this paper. For the movement phase, the particles move following the dynamics of Gaussian process or ballistic type of L\'evy walk, and the time of each movement is random. For the return phase, the particles will move back to the origin with a constant velocity or acceleration or under the action of a harmonic force after each movement, so that this phase can also be treated as a non-instantaneous resetting. After each return, a rest with a random time at the origin follows. The asymptotic behaviors of the mean squared displacements with different kinds of movement dynamics, random resting time, and returning are discussed.
The stationary distributions are also considered when the process is localized. Besides, the mean first passage time is considered when the dynamic of movement phase is Brownian motion.

\end{abstract}

\maketitle

\section{Introduction}

Resets are needed or discovered in many different fields, such as ecology \cite{benichou2005,Lomholt11055,Loverdo2009}, biology \cite{dgar,Srikanth}, and computer science \cite{Andrea}. One of the most representative examples of reset in ecology is intermittent search processes for the foraging animal, which needs to return home from time to time. In \cite{Martin}, the authors also illustrate that the diffusions with reset are able to capture the global dynamics through the different states of the animals. In daily life, returning home after finishing work can also be considered as a phenomenon of reset. Resetting means that a process returns to a given position and starts its motion anew from time to time. Resetting is also a natural strategy when one gets on the wrong track while trying to find a solution \cite{Eule}.

The Brownian motion and L\'evy walk can be regarded as two representative categories of diffusions which can be distinguished by the relationship between the mean squared displacement (MSD) $\langle x^2(t)\rangle$ and time $t$. Specifically, $\langle x^2(t)\rangle\sim t^\alpha$ with $\alpha=1$ for normal diffusion such as Brownian motion \cite{NelsonE}, while the process with $\alpha \neq 1$ is called anomalous diffusion \cite{BG1990,SLLS2004,MJCB2014}, e.g., L\'evy walk. Two popular and well-developed models to describe anomalous diffusions are continuous time random walk (CTRW) with independent space and time \cite{s11,First} and the L\'evy walks \cite{Zaburdaev,First}.  For the CTRW model, the two categories of independent identically distributed (i.i.d.) random variables known as waiting time and jump length are independent with each other \cite{s16}, while the dynamic of L\'evy walk is space-time coupled through finite propagation speed \cite{Zaburdaev}. Recently, the methods of Hermite polynomial expansions are established to solve some issues of L\'evy walk \cite{XDS2020}, essentially being effective for L\'evy walk with external potential \cite{Bo,ZXD2021}.  Both CTRW and L\'evy walk have many applications in finance \cite{s3}, ecology \cite{s4}, and biology \cite{s6}.

%

The diffusion process with rests have been analyzed from various perspectives, including the different mechanisms of resets \cite{Justin,Miquel,Xavier}, completion time of the processes with resets \cite{Arnab,Arnab16,Chechkin}, and using resets to connect different realizations of processes  \cite{Miquel2016,Miquel2017}. Moreover, the non-equilibrium steady state as well as the mean first passage time (MFPT) have been discussed in \cite{ukasz,Arnab,Arnab16}. It is also natural to consider particle rest for a period of time to recharge after finishing each movement and returning to the origin. 
For the related discussions, see \cite{Puigdel,Martin2018,Axel2019}.
Besides, CTRW and L\'{e}vy walks with stochastic resetting have been studied in \cite{Martin,Tian}. Specifically, the particle may reset to a given position $x_r$ with a resetting rate $r\in [0,1]$ after each step of moving. The results in \cite{Tian} indicate that the resetting can make the particles localized.



In \cite{Tian,Puigdel,Martin2018}, the authors treat resetting as an instantaneous action which means return to a given point without costing of time. Besides in \cite{Axel}, a two-state model has been developed, where the two states respectively correspond to the phases of movement and returning to the origin at a constant velocity. Therefore, it takes some time for the particle to reset or return to the origin after each movement phase, which also indicates non-instantaneous reset, and some statistical quantities, such as MSDs, are calculated. In \cite{Anna}, more ways of returning, such as returning at constant acceleration and under action of harmonic force, are considered for the two-state model where the dynamic of movement phase is Gaussian.



In this paper, we focus on the process with three different phases, i.e., movement phase, return phase, and rest phase. For the first phase, Brownian motion and ballistic type of L\'evy walk are considered, respectively. As for the second phase, the types of returning are the same as the ones in \cite{Anna}, which are returning with a constant speed, acceleration or return in the harmonic potential. Additionally, it is natural to consider the rest for a stochastic period of time  after each moving and returning.
%
Another difference between \cite{Anna} and this paper is that we do not require the first moment of every period to be finite. Besides, we not only consider the dynamic of movement phase to be Gaussian process discussed in  \cite{Anna,Axel}, but also discuss L\'evy walk type of moving in first phase, especially the ballistic type of L\'evy walks. The Lamperti distribution \cite{Barkai,Froemberg} which describes the asymptotic density of L\'{e}vy walk in the regime of ballistic scaling plays a key role in our discussion.
%

This paper is organized as follows. In Section \ref{sec 2}, we introduce the three-state model, and generally analyze the asymptotic behaviors of MSDs for different types of returns when time $t$ is long enough; the probability density function (PDF) of return time and its moments are also discussed for each case. In Section \ref{sec 3},  we turn to discuss the case that the propagator function behaves as Gaussian distribution for different types of PDFs of movement time and rest time regarding to each kind of return; some transport properties are considered. Next in Section \ref{sec 4}, the asymptotic behavior of the particle is analyzed when the propagator function is obtained from the ballistic type of L\'{e}vy walk. Finally, the first arrival statistics are considered in Section \ref{sec 5} when the propagator function is Gaussian distribution.

%
%
%

\section{Model}\label{sec 2}

The model discussed in this paper is in one dimensional space, consisting of three different phases, namely, movement phase, return phase, and rest phase. 
 The stochastic process $X_t$ may have several cycles up to time $t$,  
and is assumed to start at the origin with the movement phase. Denote $\{\eta_i\}$ as independent and identically distributed (i.i.d.) positive random variables representing the durations of each movement phase, and $\psi_m(\eta)$ as the corresponding PDF of each $\eta_i$. Besides, the position of the $i$th movement phase $M_\tau$ with $0<\tau\leq \eta_i$ follows a propagator density function $h(x,t)=\langle \delta (M_\tau - x) \rangle$, where $\delta(\cdot)$ represents the Dirac $\delta$ function. After each movement phase, it follows a return phase unless the process is stopped in the movement phase. In the return phase, the process will return to the origin from the end position of the previous movement phase $M_{\eta_i}$ in the way of $r = R(\tau; M_{\eta_i})$, where $R$ is a prescribed function, with a little abuse of notation $\tau$ represents the elapsed time since the beginning of each return phase, and $r$ is the position after returning for a period of time $\tau$. Besides we denote $\rho_i$ as the time spent by the process to return to the origin, i.e., $0=R(\rho_i; M_{\eta_i})$, and obviously $0<\tau\leq\rho_i$. 
After the return phase, this cycle has not yet finished; it will come to the phase of rest, which means the process will stay at the origin for a period of time $\{\xi_i\}$.
Here we also assume that $\{\xi_i\}$ are i.i.d. random variables following the density function $\phi_r(\xi)$. 
After the rest phase, the process will start a new cycle.
Denote $\Theta_{n}=\sum_{i=0}^n (\eta_i + \rho_i + \xi_i)$ and $N_t = \max\{n; \Theta_n \leq t \}$. Then the position of the process with the three phases at time $t$ can be given as
\begin{equation*}
	X_t = \left\{
	\begin{split}
		 M_{t-\Theta_{N_t}},  \quad &\text{if}~ \Theta_{N_t}+\eta_{N_t+1} > t;\\
		 r_{N_t+1}, \quad & \text{if}~ \Theta_{N_t}+\eta_{N_t+1} \leq t\\
		& \quad \text{and}~ \Theta_{N_t}+\eta_{N_t+1} + \rho_{N_t+1} > t;\\
		 0, \quad  & \text{if}~ \Theta_{N_t}+\eta_{N_t+1} + \rho_{N_t+1} \leq t,
	\end{split}
	\right.
\end{equation*}
where $r_{N_t+1} = R\big(t-(\Theta_{N_t} + \eta_{N_t+1}); M_{\eta_{N_t+1}}\big)$. For the convenience of discussion, we additionally assume that the propagator density function $h(x,t)$ is even, i.e., $h(x,t)=h(-x,t)$ and the return function satisfies $R(\tau; M_{\eta_i})=-R(\tau; -M_{\eta_i})$.

In order to obtain the PDF $P(x,t)$ of process $X_t$ defined as $P(x,t) = \langle \delta(X_t - x)\rangle$, we first discuss the probabilities of just starting the movement phase at time $t$, the return phase at time $t$ and position $x$, the rest phase at time $t$, and denote them as $j_1(t), j_2(x,t), j_3(t)$, respectively.
%
According to the model described above, it holds that
\begin{equation}\label{1.1}
j_1(t)=\delta(t)+\int_{0}^{t}j_3(t-t')\phi_r(t')dt',
\end{equation}
where the first term on the right hand side (r.h.s.) means that the process starts with movement phase and the second term represents that the particle has just arrived at rest phase at time $(t-t')$ and keeps resting for time $t'$ until it starts to move again. Besides the following relation between $j_1(t)$ and $j_2(x,t)$ also holds
\begin{equation}\label{1.2}
  j_2(x,t)=\int_{0}^{t}j_1(t-t')\psi_m(t')h(x,t')dt',
\end{equation}
where the term on r.h.s. means that the particle has just arrived at movement phase at time $(t-t')$ and then moves to position $x$ according to the propagator function $h(x,t')$ and the moving time $t'$ drawn from the PDF $\psi_m(t')$. Before we give the relation between $j_2$ and $j_3$, we denote the inverse function of $r = R(\tau; x)$ as $\tau = R^{-1}(r;x)$ which means the time returning from end position of previous movement phase $x$ to position $r$. Therefore we have
\begin{equation}\label{1.3}
  j_3(t)=\int_{-\infty}^{\infty} \int_{0}^{t} j_2(x,t-t') \delta \left(t'-R^{-1}(0;x)\right) dt' dx,
\end{equation}
where the term on r.h.s. describes that the particle has just reached the return phase at position $x$ at time $(t-t')$, then it takes time $t'$ identical with $R^{-1}(0,x)$ to move from $x$ to the origin. Applying Laplace transform defined as $\hat{g}(s)=\mathscr{L}_{t\rightarrow s}\{g(t)\}=\int_{0}^{\infty} e^{-s t}g(t)dt$ to \eqref{1.1}, \eqref{1.2}, and \eqref{1.3}, we have 
\begin{align}
  \hat{j}_1(s) & =1+ \hat{j}_3(s)\hat{\phi}_r(s);\nonumber \\
  \hat{j}_2(x,s) & =\hat{j}_1(s) \mathscr{L}_{t\rightarrow s}\big\{\psi_m(t)h(x,t)\big\};\nonumber\\
  \hat{j}_3(s) & =\int_{-\infty}^{\infty}\hat{j}_2(x,s)\exp\left(-s R^{-1}(0;x)\right)dx. \nonumber
\end{align}
The expressions of $\hat{j}_1(s)$, $\hat{j}_2(x,s)$, and $\hat{j}_3(s)$ can be obtained by  solving the above equations
%
\begin{align}
& \hat{j}_1(s)   =\frac{1}{1-\hat{\phi}_r(s) \Omega(s)};\nonumber \\
 &\hat{j}_2(x,s)  =\frac{\mathscr{L}_{t\rightarrow s}\big[\psi_m(t)h(x,t)\big]}{1-\hat{\phi}_r(s)\Omega(s)}; \nonumber \\
  &\hat{j}_3(s)  =\frac{\Omega(s)}{1-\hat{\phi}_r(s)\Omega(s)},\label{j}
\end{align}
where $\Omega(s) = \int_{-\infty}^{\infty}\mathscr{L}_{t\rightarrow s}\big\{\psi_m(t)h(x,t)\big\}e^{-s R^{-1}(0;x)}dx$.

Now, we start to derive the PDF $P(x,t)$ of the process $X_t$, and the movement phase, return phase, and rest phase are respectively indexed as phase 1, 2, and 3. Besides denote the $P_i(x,t)$, $i=1, 2, 3$ as the PDF of the particle arriving at position $x$ at time $t$ in phase $i$. Obviously, we have
\begin{equation}\label{P_sum_Pi}
	P(x,t)=P_1(x,t)+P_2(x,t)+P_3(x,t).	
\end{equation}
With the help of $j_1(t)$, $j_2(x,t)$, and $j_3(t)$, the expression for each $P_i(x,t)$ can be respectively obtained as
\begin{align}
  P_1(x,t) &=\int_{0}^{t}j_1(t-t')\Psi_m(t')h(x,t')dt';\label{1.10} \\
  P_2(x,t) & =\int_{-\infty}^{\infty}\int_{0}^{t}j_2(z,t-t')\delta\big(x-R(t';z)\big)dt' dz;\label{1.11}\\
  P_3(x,t) & =\delta(x)\int_{0}^{t}j_3(t-t')\Phi_r(t')dt',\label{1.12}
\end{align}
 where $\Psi_m(t)=\int_{t}^{\infty}\psi_m(\tau)d\tau$ and $\Phi_r(t)=\int_{t}^{\infty}\phi_r(\tau)d\tau$ are the survival probabilities of moving time and resting time PDFs, respectively. In Laplace space the survival probabilities can be easily obtained by
 \begin{equation*}
 \widehat{\Psi}_m(s)=\frac{1-\hat{\psi}_m(s)}{s},\quad \widehat{\Phi}_r(s)=\frac{1-\hat{\phi}_r(s)}{s}.
 \end{equation*}
It should be noted that \eqref{1.10} represents particle just comes to the movement phase at time $(t-t')$, and stays in this phase for time $t'$, and the position is drawn from $h(x,t')$. Similarly, \eqref{1.11} indicates that the return phase begins at time $(t-t')$ from position $z$, and the particle returns from $z$ towards origin for time $t'$ then arrives at $x=R(t';z)$. Finally, \eqref{1.12} represents that the particle comes to rest phase at time $(t-t')$ and stays in this phase for the rest of time, and $\delta(x)$ indicates the position where particle rests is the origin.
Applying Laplace transform on \eqref{1.10}, \eqref{1.11}, and \eqref{1.12}, respectively, we have
%
%
 \begin{align}
   \widehat{P}_1(x,s) & =\hat{j}_1(s) \mathscr{L}_{t\rightarrow s}\left\{\Psi_m(t)h(x,t)\right\};\nonumber\\
   \widehat{P}_2(x,s) & = \int_{|x|}^{\infty}\hat{j}_2(z,s)\mathscr{L}_{t\rightarrow s}\left\{\delta\left(|x|-R(t;z)\right)\right\}dz; \nonumber \\
   \widehat{P}_3(x,s) & =\delta(x) \hat{j}_3(s)\widehat{\Phi}_r(s), \label{P}
 \end{align}
 where the assumption of $R(t;z)=-R(t;-z)$ is used in the derivation of $\widehat{P}_2(x,s)$. Finally 
 $P(x,t)$ can be obtained through \eqref{P_sum_Pi}.

%

The PDF of time returning from the final position of previous movement phase to the origin denoted as $\varphi(t)$ can be given by
%
%
 \begin{equation}\label{1.17}
   \varphi(t)= \int_{-\infty}^{\infty}\delta\left(t-R^{-1}(0;x)\right) \int_{0}^{\infty}\psi_m(t') h(x,t') dt' dx.
 \end{equation}
 The r.h.s. of \eqref{1.17} means that the particle has arrived at position $x$ at the end of the movement phase, and takes time $R^{-1}(0;x)$ to return to the origin from $x$. It is natural that rest phase has no influence on the return time, which can also be seen from \eqref{1.17}.
%

Next, we will consider several representative methods of returns. The PDF $P(x,t)$ and return time PDF $\varphi(t)$ are obtained; moreover, some of the properties are also discussed so that we can unravel the influences of different types of rests.

\subsection{Return at a constant speed}\label{sec_2_a}

When the particle returns to origin at a constant speed $v>0$, the corresponding return dynamic behaves as
\begin{equation*}
  x=R(t;y)=
  \begin{cases}
    y - v t, & \mbox{if $y \geq 0$}; \\
    y + v t, & \mbox{otherwise}.
  \end{cases}
\end{equation*}
The corresponding inverse function of $x=R(t;y)$ is
$$t=R^{-1}(x;y)=\frac{|x-y|}{v}.$$
Therefore, the $\Omega(s)$ in \eqref{j} denoted as $\Omega_v(s)$ is given by
\begin{equation*}
	\Omega_v(s) = \int_{-\infty}^{\infty}\mathscr{L}_{t\rightarrow s}\left\{\psi_m(t)h(x,t)\right\}e^{-s \frac{|x|}{v}}dx.
\end{equation*}
Further from \eqref{j} and \eqref{P}, we have
\begin{align}
   \widehat{P}_1(x,s) & =\frac{ \mathscr{L}_{t\rightarrow s}\big\{\Psi_m(t)h(x,t)\big\}}{1-\hat{\phi}_r(s)\Omega_v(s)};\nonumber \\
   \widehat{P}_2(x,s) & = \frac{\frac{1}{v} e^{s\frac{|x|}{v}}\int_{|x|}^{\infty}\mathscr{L}_{t\rightarrow s}\big\{\psi_m(t)h(z,t)\big\}e^{-s \frac{z}{v}}dz}{1-\hat{\phi}_r(s)\Omega_v(s)}.\nonumber\\
   \widehat{P}_3(x,s) & =\frac{\delta(x)\widehat{\Phi}_r(s)\Omega_v(s)}{1-\hat{\phi}_r(s)\Omega_v(s)}.\label{1.22}
 \end{align}
 The overall PDF $P(x,t)$ can be obtained by $\widehat{P}(x,s)=\widehat{P}_1(x,s)+\widehat{P}_2(x,s)+\widehat{P}_3(x,s)$. From the definition of MSD $\langle x^2(t)\rangle$, the corresponding Laplace transform $\langle \hat{x}^2(s)\rangle$ satisfies
 \begin{equation*}
 \begin{split}
 	\langle \hat{x}^2(s)\rangle =\int_{-\infty}^{\infty}x^2 \widehat{P}(x,s)dx.
 \end{split}
 \end{equation*}
In the following, we will utilize \eqref{P_sum_Pi} and \eqref{1.22} to calculate $\langle \hat{x}^2(s)\rangle$. Obviously, $\int_{-\infty}^{\infty} x^2 \widehat{P}_3(x,s)dx=0$ because of $\delta(x)$. Therefore, there exists
 \begin{equation*}
 \begin{split}
   \langle \hat{x}^2 & (s)\rangle =\frac{\mathscr{L}_{t\rightarrow s}\big\{ \Psi_m(t)\langle x^2(t)\rangle_h \big\}}{1-\hat{\phi}_r(s) \Omega_v(s)}\\
   &+\frac{\int_{-\infty}^{\infty}x^2 e^{s\frac{|x|}{v}}\int_{|x|}^{\infty}\mathscr{L}_{t\rightarrow s}\big\{\psi_m(t) h(z,t)\big\}e^{-s\frac{z}{v}}dz dx}{v[1-\hat{\phi}_r(s) \Omega_v(s)]},
   \end{split}
 \end{equation*}
where $\langle x^2 (t)\rangle_h = \int_{-\infty}^{\infty} x^2 h(x,t)dx$. By preserving the leading term, the asymptotic form of MSD in Laplace space can be given as
\begin{equation}\label{MSDv}
\begin{split}
   \langle \hat{x}^2(s)\rangle  \sim \frac{\mathscr{L}_{t\rightarrow s}\big\{\Psi_m(t)\langle x^2(t)\rangle_h\big\}}{1-\hat{\phi}_r(s)\left[\hat{\psi}_m(s)-\frac{s}{v}\mathscr{L}_{t\rightarrow s}\big\{\psi_m(t)\langle |x(t)|\rangle_h\big\}\right]} \\
      +\frac{\frac{1}{3 v}\mathscr{L}_{t\rightarrow s}\big\{\psi_m(t)\langle |x(t)|^3\rangle_h\big\}}{1-\hat{\phi}_r(s)\left[\hat{\psi}_m(s)-\frac{s}{v}\mathscr{L}_{t\rightarrow s}\big\{ \psi_m(t)\langle |x(t)|\rangle_h\big\}\right]} 
\end{split}
\end{equation}
It can be found that the numerators of \eqref{MSDv} are consistent with the ones of \cite{Axel}, while the denominators are different reflecting the influence of rest phase. From \eqref{MSDv}, it can also be noted that the MSD w.r.t. $P(x,t)$ also depends on the first and third absolute moments and second moment w.r.t. the propagator $h(x,t)$.
%
According to \eqref{1.17} and the scaling property of Dirac $\delta$ function, in this case
the returning time PDF 
\begin{equation}\label{1.25}
   \varphi(t)= 2 v \int_{0}^{\infty}\psi_m(t') h(v t,t') dt',
\end{equation}
which is in accordance with the one of \cite{Axel}.
%
%
Then we can further obtain the $n$th moment of the return time 
\begin{equation}\label{mmt_tr}
	\langle t_r^n\rangle = \frac{1}{v^n} \int_0^\infty \psi_m(t)\big\langle |x(t)|^n \big\rangle_h dt,
\end{equation}
where the symmetry of $h(x,t)$ w.r.t. $x$ is used.

\subsection{Return at a constant acceleration}\label{sec_2_b}

In this part, we consider that the particle returns under the influence of a constant acceleration $a$, so that the force is a constant given by $F = ma$, where $m$ represents the mass of particle. For the convenience of calculation, we assume that the particle has unit mass $m=1$. Under the assumptions that the initial velocity is $0$ and the initial position of each return phase is final position $y$ of the previous movement phase, then the dynamic of particle in the return phase can be given by
\begin{equation}\label{1.27}
  \begin{cases}
    x(t)=y+a t^2/2, &  \\
    v(t)=a t, &
  \end{cases}
\end{equation}
where $a<0$ (or $a>0$) when $y>0$ (or $y<0$), the sign of $a$ represents the direction of acceleration.
%
%
According to \eqref{1.27}, the term $\Omega(s)$ in \eqref{j} denoted as $\Omega_{|a|}(s)$ in this part can be given by
\begin{equation*}
	\Omega_{|a|}(s)=\int_{-\infty}^{\infty}\mathscr{L}_{t\rightarrow s}\big\{\psi_m(t) h(x,t) \big\} e^{-s \sqrt{\frac{2 |x|}{|a|}}}dx,
\end{equation*}
which leads to $\hat{j}_1(s)$, $\hat{j}_2(x,s)$, and $\hat{j}_3(s)$ from \eqref{j}, respectively.
%
Further from \eqref{P}, there exists
\begin{align}
   \widehat{P}_1(x,s) &=\frac{ \mathscr{L}_{t\rightarrow s}\big\{ \Psi_m(t) h(x, t) \big\}}{1-\hat{\phi}_r(s) \Omega_{|a|}(s)}; \nonumber \\
   \widehat{P}_2(x,s)& = \frac{\int_{|x|}^{\infty}\mathscr{L}_{t\rightarrow s}\big\{ \psi_m(t)h(z,t)\big\}\frac{e^{-s\sqrt{\frac{2 (z-|x|)}{|a|}}}}{\sqrt{z-|x|}}dz}{\sqrt{2 |a|}\left(1-\widehat{\phi}_r(s)\Omega_{|a|}(s)\right)};\nonumber \\
   \widehat{P}_3(x,s) &=\frac{\delta(x)\widehat{\Phi}_r(s)\Omega_{|a|}(s)}{1-\hat{\phi}_r(s)\Omega_{|a|}(s)}. \label{1.29}
 \end{align}
 The overall propagator PDF $P(x,t)$ can also be obtained from  \eqref{P_sum_Pi}. Similarly, the asymptotic behavior of MSD in Laplace space w.r.t. time $t$ is obtained by keeping the leading term, which is
 \begin{equation}\label{MSDa}
 \begin{split}
    &\langle \hat{x}^2 (s)\rangle \\
    &\sim \frac{\mathscr{L}_{t\rightarrow s}\big\{\Psi_m(t) \langle x^2(t)\rangle_h \big\}}{1-\hat{\phi}_r(s) \left[\hat{\psi}_m(s) - s\sqrt{\frac{2}{|a|}}\mathscr{L}_{t\rightarrow s}\big\{\psi_m(t)\langle |x|^{\frac{1}{2}}(t)\rangle_h\big\} \right]} \\
      & +\frac{\frac{8}{15}\sqrt{\frac{2}{|a|}}\mathscr{L}_{t\rightarrow s}\big\{ \psi_m(t) \langle |x|^{\frac{5}{2}}(t)\rangle_h \big\}}{1 -\hat{\phi}_r(s)\left[\hat{\psi}_m(s)-s\sqrt{\frac{2}{|a|}}\mathscr{L}_{t\rightarrow s}\big\{\psi_m(t) \langle |x|^{\frac{1}{2}}(t)\rangle_h\big\}\right]}.
 \end{split}
 \end{equation}
Therefore, the overall MSD is influenced not only by the PDFs of movement time and rest time, the constant acceleration, but also by moments $\langle |x|^{\frac{1}{2}}(t)\rangle_h$, $\langle |x|^{\frac{5}{2}}(t)\rangle_h$, and $\langle x^{2}(t)\rangle_h$.
%
Further the expression of return time PDF $\varphi(t)$ can be obtained from \eqref{1.17}, which is
\begin{equation}\label{1.30}
   \varphi(t)= 2 |a| t \int_{0}^{\infty}\psi_m(t') h\left(\frac{|a|}{2}t^2, t'\right) dt'.
 \end{equation}
The $n$th moment of return time $\langle t_r^n\rangle$ can be obtained by multiplying $t^n$ and integrating over $t>0$ on both sides of \eqref{1.30}. Then we have
  \begin{equation}\label{1.31}
   \langle t_r^n\rangle=\left (\frac{2}{|a|}\right)^{\frac{n}{2}}\int_{0}^{\infty} \psi_m(t)\langle |x(t)|^{\frac{n}{2}} \rangle_h dt.
 \end{equation}
It can be observed from \eqref{1.31} that the $n$th moment of return time depends explicitly on movement time PDF $\psi_m$, the acceleration, and the $(n/2)$th absolute moment w.r.t. $h(x,t)$.

\subsection{Influence of harmonic force on return phase}\label{sec_2_c}

In this part, we still assume that the particle is with unit mass, i.e., $m=1$, while it is under the influence of a harmonic potential $U(x)=kx^2/2$, $k>0$  during the return phase.
Besides, the initial speed of each return phase is assumed to be zero, and it is natural to let the initial position be the final point of the previous movement phase $y$. Therefore, the dynamic of return phase is
%
\begin{equation*}
    x(t)=y\cos{\left (\sqrt{k} t \right)}.
\end{equation*}
It can be found that $R^{-1}(0;y)=\frac{\pi}{2\sqrt{k}}$. Then we have
\begin{equation*}
	\begin{split}
		\Omega(s) & = \int_{-\infty}^{\infty}\mathscr{L}_{t\rightarrow s}\big\{\psi_m(t)h(y,t)\big\}e^{-s \frac{\pi}{2\sqrt{k}}} dy\\
		& = \hat{\psi}_m(s) e^{-s \frac{\pi}{2\sqrt{k}}},
	\end{split}
\end{equation*}
where the second equation utilizes normalization of $h(x,t)$. Similarly, we have
\begin{align}
 \hat{j}_1(s)  & =\frac{1}{1-\hat{\phi}_r(s)\hat{\psi}_m(s)e^{-s \frac{\pi}{2\sqrt{k}}}}; \nonumber \\
 \hat{j}_2(x,s) & =\frac{\mathscr{L}_{t\rightarrow s}\big\{ \psi_m(t) h(x,t)\big\}}{1-\hat{\phi}_r(s)\hat{\psi}_m(s) e^{-s \frac{\pi}{2\sqrt{k}}}}; \nonumber\\
 \hat{j}_3(s) & =\frac{\hat{\psi}_m(s) e^{-s \frac{\pi}{2\sqrt{k}}}}{1-\hat{\phi}_r(s)\hat{\psi}_m(s) e^{-s \frac{\pi}{2\sqrt{k}}}};  \nonumber
\end{align}
and
\begin{align}
   \widehat{P}_1(x,s) & =\frac{ \mathscr{L}_{t\rightarrow s}\big\{ \Psi_m(t)h(x,t)\big\}}{1 - \hat{\phi}_r(s)\hat{\psi}_m(s)e^{-s \frac{\pi}{2\sqrt{k}}}}; \nonumber\\
   \widehat{P}_2(x,s) & = \frac{ \int_{|x|}^{\infty}\mathscr{L}_{t\rightarrow s}\big\{ \psi_m(t)h(z,t)\big\} \frac{e^{-\frac{s}{\sqrt{k}}\arccos\left(\frac{|x|}{z}\right)}}{\sqrt{k(z^2-x^2)}}dz}{1-\hat{\phi}_r(s)\hat{\psi}_m(s)e^{-s \frac{\pi}{2\sqrt{k}}}}; \nonumber \\
   \widehat{P}_3(x,s) & =\frac{\delta(x)\hat{\psi}_m(s)\widehat{\Phi}_r(s)e^{-s\frac{\pi}{2\sqrt{k}}}}{1-\hat{\phi}_r(s)\hat{\psi}_m(s)e^{-s \frac{\pi}{2\sqrt{k}}}}. \label{1.34}
 \end{align}
Then, the overall PDF $\widehat{P}(x,s)$ can be obtained by summing up $\widehat{P}_1(x,s),\widehat{P}_2(x,s)$, and $\widehat{P}_3(x,s)$. The corresponding asymptotic behavior of MSD in Laplace space for long time $t$ can be got as
 \begin{equation}\label{MSDk}
 \begin{split}
    &\langle \hat{x}^2 (s)\rangle\\
     & \sim\frac{\mathscr{L}_{t\rightarrow s}\big\{\Psi_m(t) \langle x^2(t)\rangle_h \big\} + \frac{\pi}{4\sqrt{k}} \mathscr{L}_{t\rightarrow s}\big\{\psi_m(t) \langle x^2(t)\rangle_h \big\}}{1-\hat{\phi}_r(s)\hat{\psi}_m(s)e^{-s \frac{\pi}{2\sqrt{k}}}}.
 \end{split}
 \end{equation}
 Therefore the overall MSD is decided by $\langle x^2(t)\rangle_h$, i.e., the MSD w.r.t. propagator density $h(x,t)$,  movement and rest densities, and parameter of harmonic potential $k$.
%

\begin{figure}[htbp]
  \centering
  \includegraphics[width=8cm]{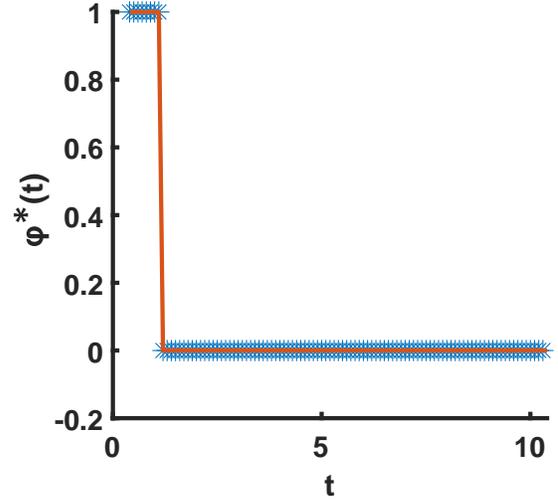}
  \caption{Survival probability of return time PDF $\varphi^*(t)$ for the case of harmonic force. The propagator density $h(x,t)= \frac{1}{\sqrt{4\pi D t}}e^{-x^2/(4 D t)}$  with $D=5$, movement PDF  $\psi_m(t) = \alpha e^{-\alpha t}$ with $\alpha = 0.75$, and parameter $k=2$. The simulation results are obtained by averaging over $5\times10^4$ realizations.}
  \label{surk}
\end{figure}

The return time PDF $\varphi(t)$ can also be obtained from \eqref{1.17}
  \begin{equation}\label{1.35}
   \varphi(t)= \delta\left(t-\frac{\pi}{2\sqrt{k}}\right).
 \end{equation}
 In order to verify the result in \eqref{1.35} by numerical simulations, equivalently we can calculate the corresponding survival probability of $\varphi(t)$, which is given by
 \begin{equation*}
 \begin{split}
 	\varphi^*(t) &= \int_t^\infty \varphi(\tau)d\tau = \theta\left(\frac{\pi}{2 \sqrt{k}} - t \right).
 \end{split}
 \end{equation*}
 The corresponding numerical simulations are presented in  Fig. \ref{surk}.
%
Further the $n$th moment of return time $\langle t_r^n\rangle$ can be directly obtained from \eqref{1.35}, i.e.,
 \begin{equation}\label{1.36}
   \langle t_r^n\rangle=\left(\frac{\pi}{2\sqrt{k}}\right)^n.
 \end{equation}

Different from the first two cases, when the process with three phases is influenced by a harmonic force on return phase, the corresponding return time PDF $\varphi(t)$ and $n$th moment $\langle t_r^n\rangle$ have no relevance to rest PDF $\phi_r(t)$ or propagator density $h(x,t)$. They are totally decided by harmonic potential.

\section{
Particles with Gaussian propagator in the movement phase: MSD and return time PDF
}\label{sec 3}

In this section we assume that the propagator density function of the particle in movement phase behaves as Gaussian distribution $h(x,t)=\frac{1}{\sqrt{4 \pi D t}} e^{-\frac{x^2}{4 D t}}$. In the following we can observe the influences of rest by calculating the representative statistical quantities and distributions, such as MSDs, return time PDFs, and stationary distributions.
%

\subsection{Exponentially distributed movement time}

In this part, we consider the PDF for the time of each movement phase is in exponential form, i.e., $\psi_m(t)=\alpha e^{-\alpha t}$, $\alpha>0$, which implies the memoryless for the time of each movement phase, i.e., the Markovian property.
%
The corresponding survival probability is $\Psi_m(t)=e^{-\alpha t}$. In the following we will focus on the different ways of return discussed in Section \ref{sec 2}.

\subsubsection{Return at a constant speed}

First we consider the return time PDF $\varphi(t)$ and the $n$th moment $\langle t_r^n\rangle$ of the return time $t_r$ for the particle returning at a constant speed $v$. Starting from \eqref{1.25} and substituting $h(x,t)$ with Gaussian distribution $h(x,t)=\frac{1}{\sqrt{4 \pi D t}} e^{-\frac{x^2}{4 D t}}$ yield
\begin{equation*}
  \varphi(t)=v \sqrt{\frac{\alpha}{D}} e^{-\sqrt{\frac{\alpha}{D}}v t}.
\end{equation*}
Further from \eqref{mmt_tr} we have
\begin{equation*}
  \langle t_r^n\rangle=\frac{1}{v^n}\bigg(\frac{D}{\alpha}\bigg)^{\frac{n}{2}}\Gamma(1+n).
\end{equation*}
These results are the same as the ones in \cite{Axel}.

\begin{figure}[htbp]
  \centering
  \includegraphics[scale=0.28]{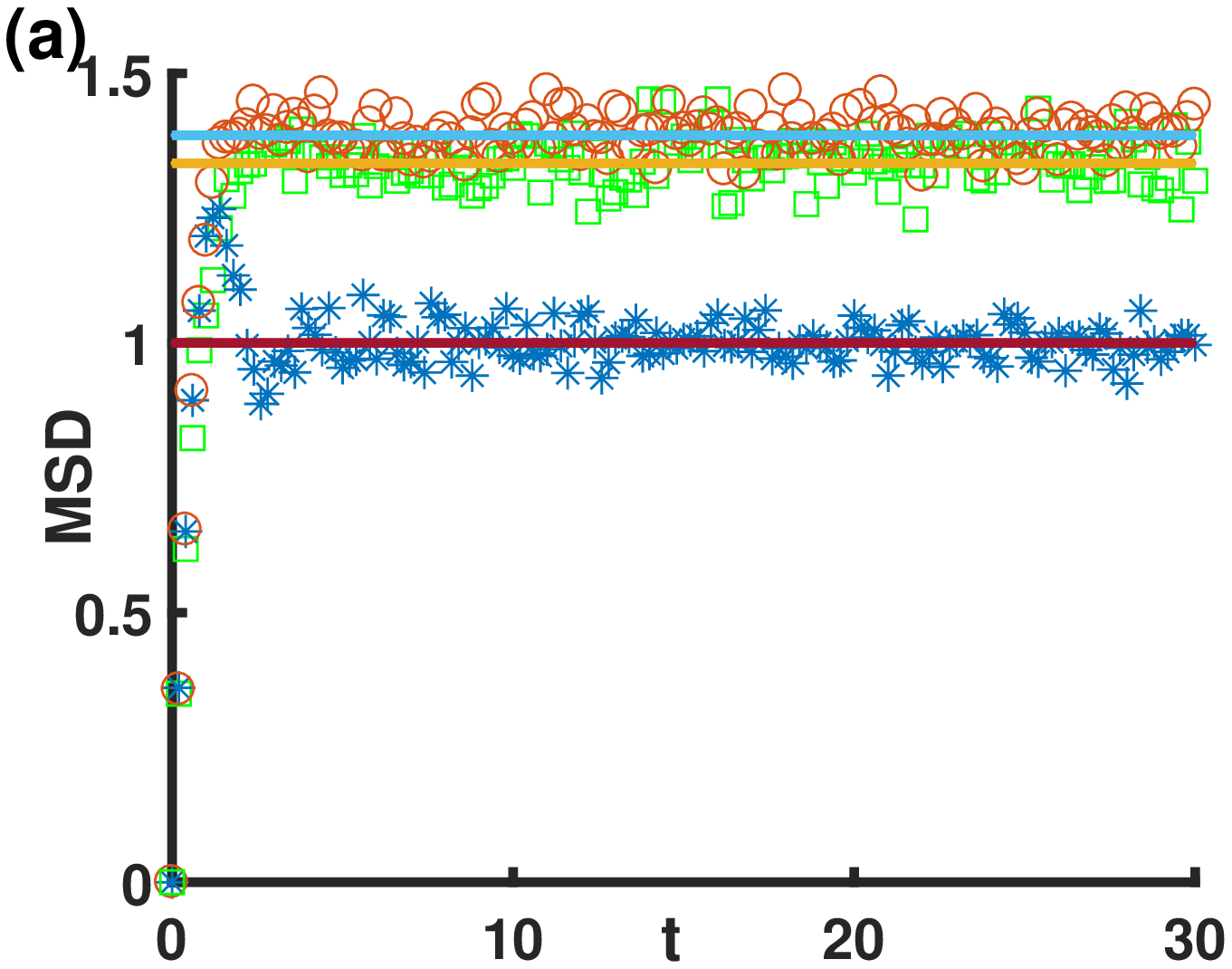}
  \includegraphics[scale=0.28]{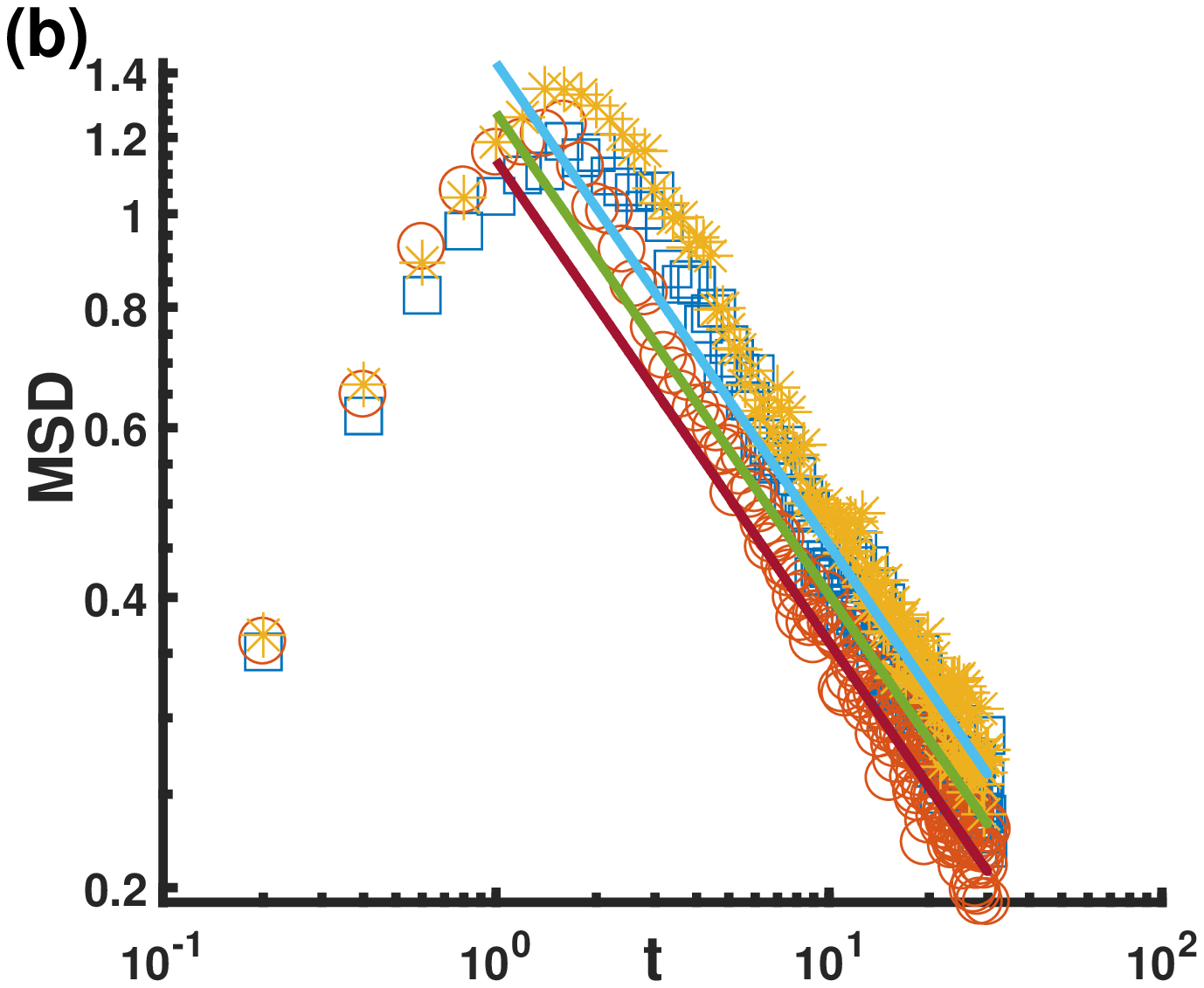}
  \caption{The overall MSDs of the particles with two types of rests, respectively, returning at constant velocity, acceleration, and under harmonic force.
  The propagator function is Gaussian and the PDF of movement time is assumed to be exponential with $\alpha=1$; the speed, acceleration, and parameter of harmonic force are $v=1$ (squares), $|a|=1$ (stars), and $k=1$ (circles), respectively. For (a), the rest time PDF is exponential distribution with $\lambda=1$. While in (b) (log-log scale), the PDF of rest time is assumed to be power-law distribution with $\beta=0.5$. The simulation results are obtained by averaging over $10^4$ realizations.}
  \label{msdgaussianmarkovie}
\end{figure}

Next we consider the overall MSD of the three-phase process with each rest time being exponentially distributed $\phi_r(t)=\lambda e^{-\lambda t}$. Then from \eqref{MSDv} we have
\begin{equation*}
  \langle \hat{x}^2 (s)\rangle \sim \frac{2 D}{\alpha}\frac{1+\frac{\sqrt{D \alpha}}{v}}{s+\frac{s \sqrt{D \alpha}}{v}+\frac{s \alpha}{\lambda}};
\end{equation*}
after applying the inverse Laplace transform, it has
\begin{equation}\label{2.4}
  \langle x^2 (t)\rangle \sim \frac{2 D}{\alpha}\frac{1+\frac{\sqrt{D \alpha}}{v}}{1+\frac{ \sqrt{D \alpha}}{v}+\frac{\alpha}{\lambda}},
\end{equation}
which is verified by the numerical simulations shown in Fig. \ref{msdgaussianmarkovie}. The overall MSD without rest phase can be found in \cite{Axel}, which is $\langle x^2 (t)\rangle\sim \frac{2 D}{\alpha}$. By comparing these two results, one can find that the rest phase makes the MSD smaller, and the return velocity $v$ also makes an effect on the MSD, being different from the case of no rest. Specifically, it can be found from \eqref{2.4} that the overall MSD $\langle x^2 (t)\rangle$ decreases with $v$, which is reasonable since for a given running time $t$ the bigger velocity $v$ is, the time of each return phase will be less so that the number of rest phases increases, which leads to small $\langle x^2 (t)\rangle$. Besides,  the results in \cite{Puigdel} with instantaneous reset can be recovered by taking $v\rightarrow\infty$.

\begin{figure}[htbp]
  \centering
  \includegraphics[width=8cm]{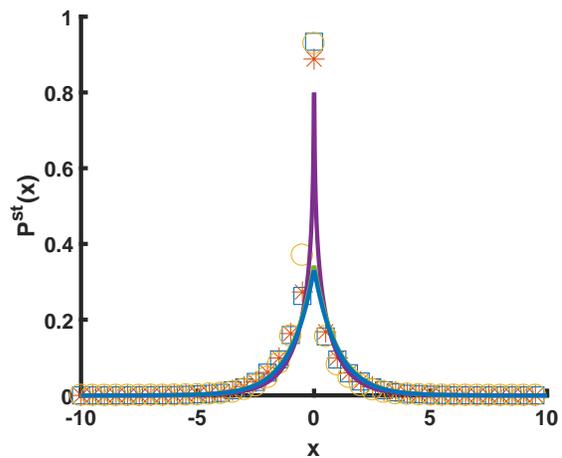}
  \caption{Numerical simulations of the stationary density $P^{st}(x)$ for different types of return. Propagator distribution is Gaussian, and the PDFs of movement time and rest time are assumed to be exponential distribution with $\alpha=\lambda=1$. The other parameters are $v=1$ (squares), $|a|=1$ (stars), and $k=1$ (circles). The simulation results are obtained by sampling over $5\times10^4$ realizations.}
  \label{pst}
\end{figure}


Moreover, the stationary distribution $P_{s t}(x)=\lim_{t \to \infty} P(x,t)$ can be obtained according to the final value theorem of Laplace transform, $P_{s t}(x)=\lim_{s\rightarrow 0} s\widehat{P}(x,s)$. With the help of \eqref{1.22} and \eqref{P_sum_Pi}, we have
\begin{equation*}
  P_{s t}(x)=\frac{\frac{1}{2\sqrt{D \alpha}}e^{-\sqrt{\frac{\alpha}{D}}|x|}+\frac{1}{2 v}e^{-\sqrt{\frac{\alpha}{D}}|x|}+\frac{\delta(x)}{\lambda}}{\frac{1}{\alpha}+\frac{1}{v}\sqrt{\frac{D}{\alpha}}+\frac{1}{\lambda}},
\end{equation*}
which is verified by the simulations (see Fig. \ref{pst}).
%

 Next, we choose the PDF of rest time as a power-law distribution, i.e.,
 \begin{equation}\label{pw-law_PDF}
 	\phi_r(t)=\frac{1}{\tau_0}\frac{\beta}{(1+t/\tau_0)^{1+\beta}},\quad 0<\beta<1,~\tau_0>0.
 \end{equation}
Substituting it into \eqref{MSDv} and taking inverse Laplace transform yield
\begin{equation}\label{2.6}
  \langle x^2 (t)\rangle \sim \frac{\sin(\pi \beta)}{\pi}\frac{2 D v+2 \alpha^{\frac{1}{2}} D^{\frac{3}{2}}}{v \alpha^2 \tau_0^\beta }t^{\beta-1}.
\end{equation}
It can be observed from \eqref{2.6} that the overall MSD decreases when time $t$ is large enough. Intuitively speaking, the process with three different phases can be considered as a competition among diffusion, return, and rest, where the latter two phases draw back the process. When the rest time follows power-law distribution, the effect of drawing back is significant when time $t$ is large enough. Besides, it can also be found that a bigger velocity will cause smaller $\langle x^2 (t) \rangle$. The result of \eqref{2.6} has been verified in Fig. \ref{msdgaussianmarkovie} through numerical simulations.

\begin{figure}[htbp]
  \centering
  \includegraphics[width=8cm]{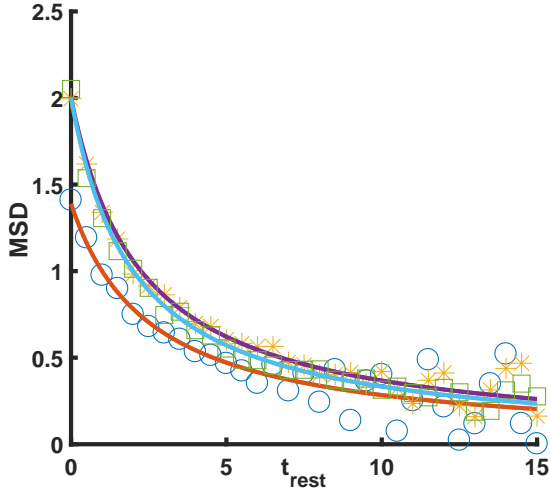}
  \caption{Numerical simulations of the MSDs for different types of return time when the particles' rest is a constant time $t_{\rm rest}$. Here, the propagator function is Gaussian distribution and the movement time follows exponential distribution with $\alpha=1$. The other parameters are $v=1$ (squares), $|a|=1$ (stars), and $k=1$ (circles). The simulation results are obtained by averaging over $5\times10^4$ realizations.}
  \label{fixrest}
\end{figure}

Furthermore, we consider the case that the rest time is a given constant $t_{\rm rest}\geq 0$, i.e., $\phi_r(t)=\delta(t-t_{\rm rest})$. Then the overall MSD is
%
\begin{equation}\label{2..4}
  \langle x^2 (t)\rangle \sim \frac{2 D}{\alpha}\frac{1+\frac{\sqrt{D \alpha}}{v}}{1+\frac{ \sqrt{D \alpha}}{v}+\alpha t_{\rm rest}}.
\end{equation}
It can be found that asymptotic behaviors of \eqref{2.4} and \eqref{2..4} are almost the same. The overall MSD $\langle x^2 (t)\rangle$ in \eqref{2..4} decreases with $t_{\rm rest}$. This result is reasonable since that the time of rest phase increases makes the proportion of movement time decrease. The $\langle x^2 (t)\rangle$ is decreasing with $v$ with the same reason as above. The result in \eqref{2..4} has been verified by simulations (see Fig. \ref{fixrest}).

\subsubsection{Return at a constant acceleration}

In this part we first consider the PDF of return time $\varphi(t)$ and the corresponding $n$th moment of return time $\langle t_r^n\rangle$ when the propagator function is still Gaussian, i.e., $h(x,t)=\frac{1}{\sqrt{4 \pi D t}} e^{-\frac{x^2}{4 D t}}$. Substituting $h(x,t)$ and $\psi_m(t)=\alpha e^{-\alpha t}$ into \eqref{1.30}, there exists
\begin{equation*}
  \varphi(t)=\frac{|a| \sqrt{\alpha}}{\sqrt{D}}t e^{-\frac{\sqrt{\alpha}|a|}{2 \sqrt{D}}t^2},
\end{equation*}
which indicates the corresponding survival probability
\begin{equation}\label{sur_prob_a_exp}
	\varphi^*(t)=e^{-\frac{\sqrt{\alpha} |a| }{2 \sqrt{D}}t^2}.
\end{equation}
The numerical simulations verify \eqref{sur_prob_a_exp}, which unravels the influences of acceleration $|a|$ and $\alpha$ on $\varphi^*(t)$. Further from \eqref{1.31}, we have
\begin{equation*}
  \langle t_r^n\rangle=\left(\frac{D}{\alpha}\right)^{\frac{n}{4}}\left(\frac{2}{|a|}\right)^{\frac{n}{2}}\Gamma\left(1+\frac{n}{2}\right).
\end{equation*}

\begin{figure}[htbp]
  \centering
  \includegraphics[scale=0.28]{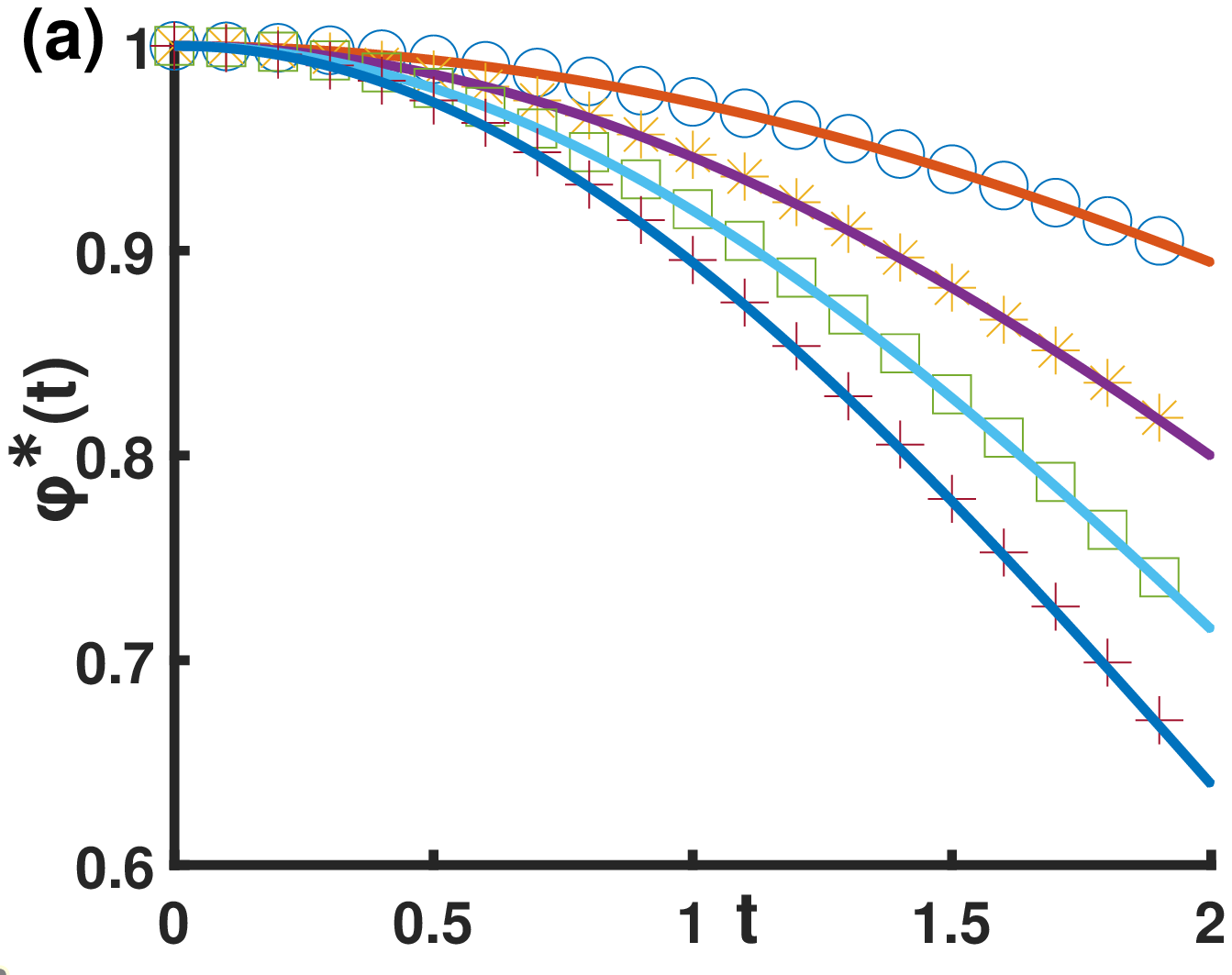}
  \includegraphics[scale=0.28]{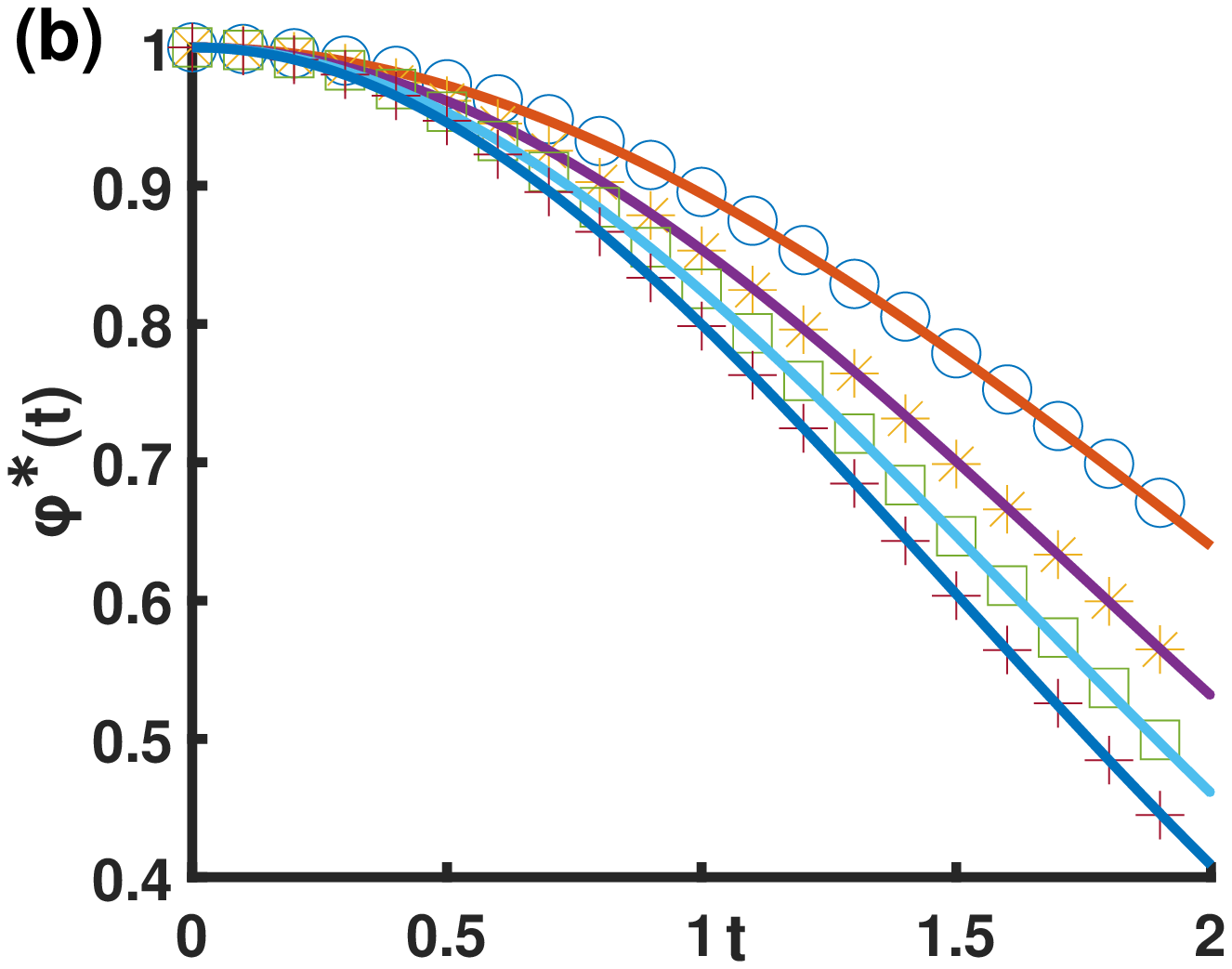}
  \caption{Survival probability $\varphi^*(t)$ of return time PDF for the particle returning at a constant acceleration $|a|$ with exponential movement time and Gaussian propagator  function. For (a), the exponent of $\psi_m$ is fixed $\alpha=0.25$, while the accelerations $|a|$, respectively, are $|a|=0.25$ (circles), $|a|=0.5$ (stars), $|a|=0.75$ (squares), and $|a|=1$ (plus). For (b), the acceleration is fixed $|a|=1$, the exponents of $\psi_m$ are $\alpha=0.25$ (circles), $\alpha=0.5$ (stars), $\alpha=0.75$ (squares), and $\alpha=1$ (plus), respectively. The simulation results are obtained by averaging over $5\times10^4$ realizations.}
  \label{suraexp}
\end{figure}

For the exponentially distributed rest time $\phi_r(t)=\lambda e^{-\lambda t}$, the corresponding asymptotic behavior of MSD can be obtained by taking inverse Laplace transform on \eqref{MSDa}, i.e.,
%
%
\begin{equation}\label{2..9}
  \langle x^2 (t)\rangle \sim \frac{2 D}{\alpha}\frac{1+\sqrt{\frac{\pi}{2|a|}}D^{\frac{1}{4}}\alpha^{\frac{3}{4}}}{1+\sqrt{\frac{\pi}{2|a|}}D^{\frac{1}{4}}\alpha^{\frac{3}{4}}+\frac{\alpha}{\lambda}},
\end{equation}
which is verified by Fig. \ref{msdgaussianmarkovie}. Similarly, the overall MSD $\langle x^2 (t)\rangle$ decreases w.r.t. the acceleration $|a|$. Roughly speaking, with the increase of $|a|$, the return time $t_r$ will be shorter so that the proportion of the rest time will increase, and it will lead to a smaller $\langle x^2 (t)\rangle$.
Moreover, the stationary distribution can be calculated with the help of \eqref{1.29} and the final value theorem of Laplace transform
\begin{equation*}
  P_{s t}(x)=\frac{\frac{1}{2\sqrt{D \alpha}}e^{-\sqrt{\frac{\alpha}{D}}|x|}+\sqrt{\frac{\pi}{8 |a|}}\Big(\frac{\alpha}{D}\Big)^{\frac{1}{4}}e^{-\sqrt{\frac{\alpha}{D}}|x|}+\frac{\delta(x)}{\lambda}}{\frac{1}{\alpha}+\Big(\frac{D}{\alpha}\Big)^{\frac{1}{4}}\sqrt{\frac{\pi}{2|a|}}+\frac{1}{\lambda}},
\end{equation*}
which is verified by the numerical simulations shown in Fig. \ref{pst}.

One can also choose the rest time following power-law distribution \eqref{pw-law_PDF}. Then from \eqref{MSDa} the corresponding inverse Laplace transform has the form
\begin{equation}\label{2.9}
  \langle x^2 (t)\rangle \sim \frac{\sin(\pi \beta)}{\pi}\frac{2 D \sqrt{|a|}+\sqrt{2 \pi} \alpha^{\frac{3}{4}} D^{\frac{5}{4}}}{\sqrt{|a|} \alpha^2 \tau_0^\beta }t^{\beta-1},
\end{equation}
where the decreasing overall MSD for sufficiently long time $t$ can also be observed. By the same methods, the asymptotic behavior of the overall MSD for the constant rest time, $\phi_r(t)=\delta(t-t_{\rm rest})$, can be given by
\begin{equation}\label{2.91}
  \langle x^2 (t)\rangle\sim \frac{2 D}{\alpha}\frac{1+\sqrt{\frac{\pi}{2|a|}}D^{\frac{1}{4}}\alpha^{\frac{3}{4}}}{1+\sqrt{\frac{\pi}{2|a|}}D^{\frac{1}{4}}\alpha^{\frac{3}{4}}+\alpha t_{\rm rest}},
\end{equation}
which has the same form as \eqref{2..9}. The theoretical result of \eqref{2.91} is confirmed by the simulations (see Fig. \ref{fixrest}).

\subsubsection{Return under the action of harmonic force}

In this part, we consider that the particle returns to the origin under the influence of harmonic force. The PDF of return time $\varphi(t)$, the corresponding survival probability $\varphi^*(t)$, and the $n$th moment of return time $\langle t_r^n\rangle$ have been discussed in \eqref{1.35} and \eqref{1.36}. Next we mainly focus on the asymptotic behaviors of overall MSDs for different types of return time while the propagator distribution $h(x,t)$ is still Gaussian, and if the MSD is a constant for sufficiently long time, then we further consider the corresponding stationary distribution.

\begin{figure}[htbp]
  \centering
  \includegraphics[scale=0.28]{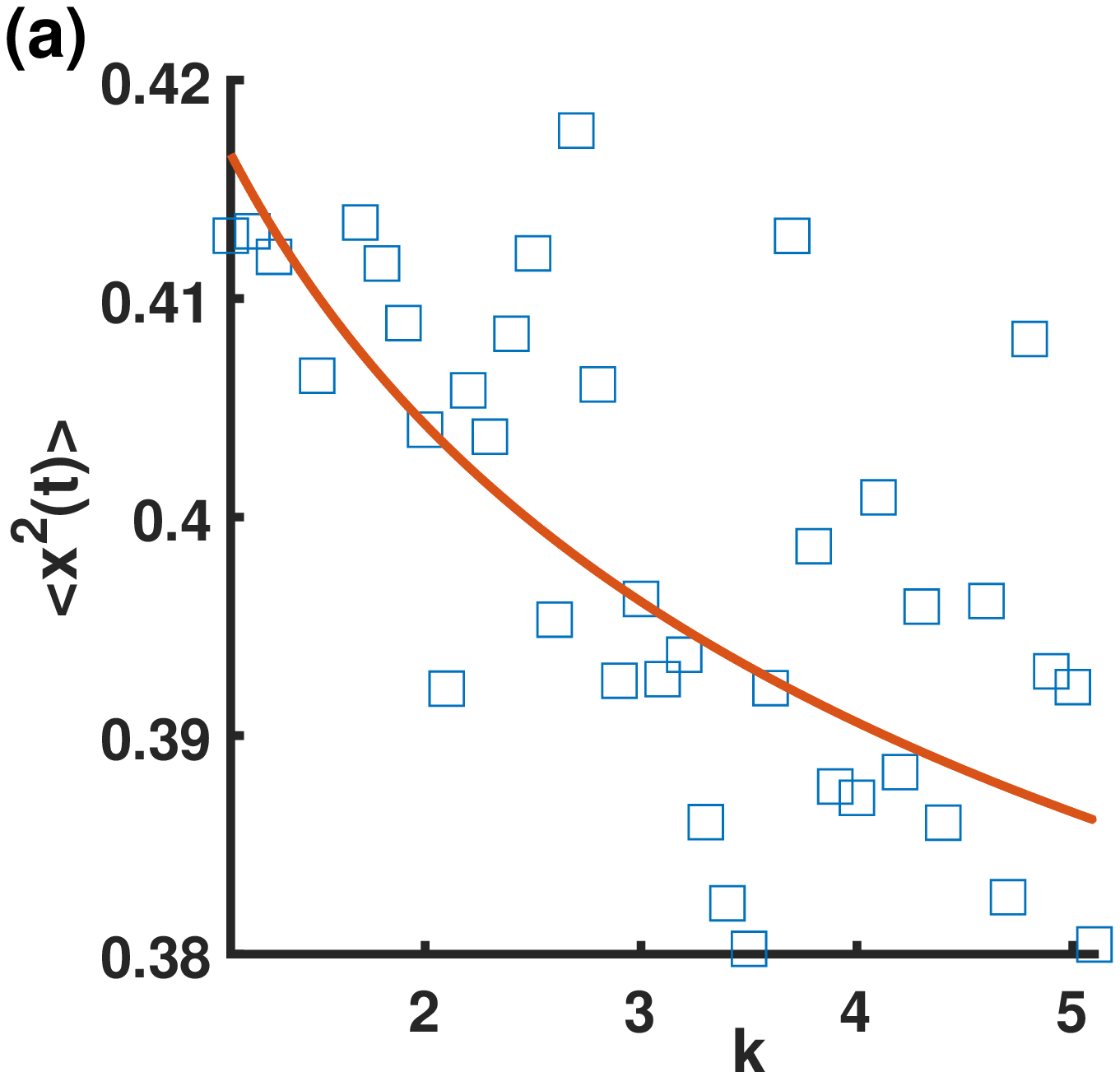}
  \includegraphics[scale=0.28]{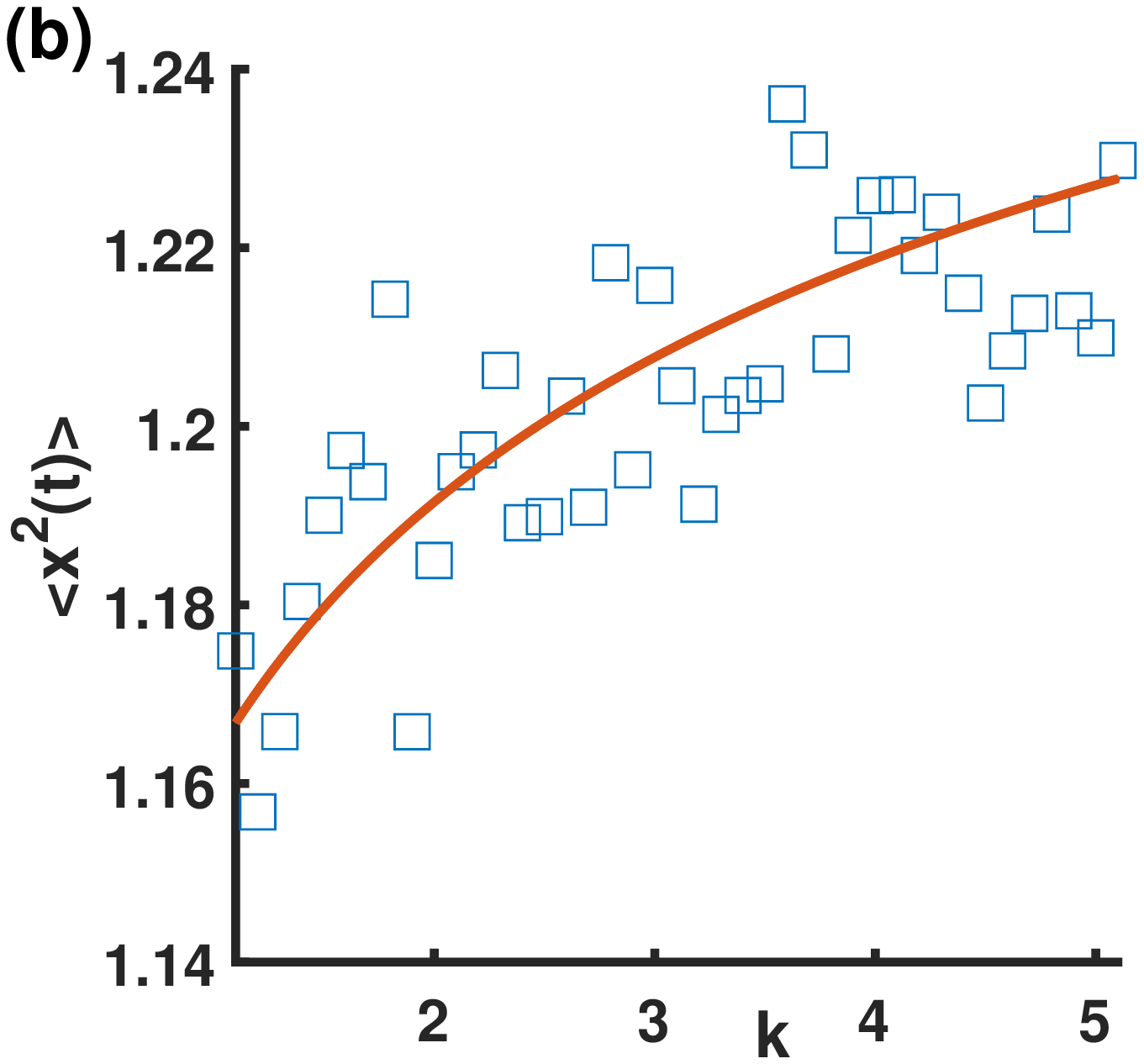}
  \caption{Numerical simulations for \eqref{MSD_BM_HF_EXP} by sampling  $5\times 10^4$ realizations. The simulation time $t=10^3$, $D=1$. For (a) $\lambda=1$ and $\alpha=2$, while for (b) $\alpha=1$ and $\lambda=2$.}
  \label{fig_MSD_BM_HF_EXP}
\end{figure}

If the rest time is exponentially distributed, then from \eqref{MSDk} we have the inverse Laplace transform
\begin{equation}\label{MSD_BM_HF_EXP}
\begin{split}
	 \langle x^2 (t)\rangle & \sim \frac{D}{\alpha}\frac{4 \sqrt{k}+ \pi \alpha}{\pi\alpha+2\sqrt{k}+2\sqrt{k}\frac{\alpha}{\lambda}}\\
	 & = \frac{D }{\alpha}\left[1-\frac{\left(\frac{\alpha}{\lambda} - 1\right)}{1 + \frac{\alpha}{\lambda}+\frac{\alpha \pi}{2 \sqrt{k}}} \right],
\end{split}
\end{equation}
which indicates that the overall MSD $\langle x^2 (t)\rangle$ is decreasing in $k$ when $\alpha>\lambda$ and increasing when $\alpha<\lambda$.
The corresponding numerical simulations are given in Fig. \ref{fig_MSD_BM_HF_EXP}. From \eqref{1.34} and final value theorem of Laplace transform, we have
\begin{equation*}
  P_{s t}(x)=\frac{\frac{1}{2\sqrt{D \alpha}}e^{-\sqrt{\frac{\alpha}{D}}|x|}+\frac{1}{2}\sqrt{\frac{\alpha}{D k}} K_0\Big(\sqrt{\frac{\alpha}{D}} |x|\Big)+\frac{\delta(x)}{\lambda}}{\frac{1}{\alpha}+\frac{\pi}{2 \sqrt{k}}+\frac{1}{\lambda}},
\end{equation*}
where $K_0(x)$ is the modified Bessel functions of the second kind. When the rest time follows power-law distribution, we have
\begin{equation*}
  \langle x^2 (t)\rangle \sim \frac{\sin(\pi \beta)}{\pi}\frac{4 D\sqrt{k}+\pi\alpha D}{2\sqrt{k} \alpha^2 \tau_0^\beta}t^{\beta-1},
\end{equation*}
which indicates the decrease of $\langle x^2 (t)\rangle$ in $t$ and $k$ for large $t$. For the deterministic time $t_{\rm rest}$, it yields
\begin{equation*}
\begin{split}
	\langle x^2 (t)\rangle &\sim \frac{D}{\alpha}\frac{4 \sqrt{k}+\pi \alpha}{ \pi\alpha+2\sqrt{k}+2\sqrt{k}\alpha t_{\rm rest}}\\
	& = \frac{D}{\alpha} \left(1-\frac{\alpha t_{\rm rest}-1}{1 + \alpha t_{\rm rest} +\frac{\pi \alpha}{2\sqrt{k}}}\right).
\end{split}
\end{equation*}
Similarly, $\langle x^2 (t)\rangle$ is decreasing in $k$ when $t_{\rm rest} > \frac{1}{\alpha}$ and increasing when $t_{\rm rest} < \frac{1}{\alpha}$.
%
%

Here we conclude that for the process with Gaussian propagator function $h(x,t)$ and exponentially distributed movement time, the process is localized when the time $t$ is large enough if the rest time follows exponential distribution or a given constant, while for the rest time following power-law distribution \eqref{pw-law_PDF}, the overall MSD decreases in time like $t^{\beta-1}$ for long time. Therefore we conjecture that if the average of rest time exists then the overall MSD is a constant, otherwise it will approximately decrease in time after long time.

\subsection{Power-law distributed movement time}

\begin{figure}[htbp]
  \centering
  \includegraphics[width=8cm]{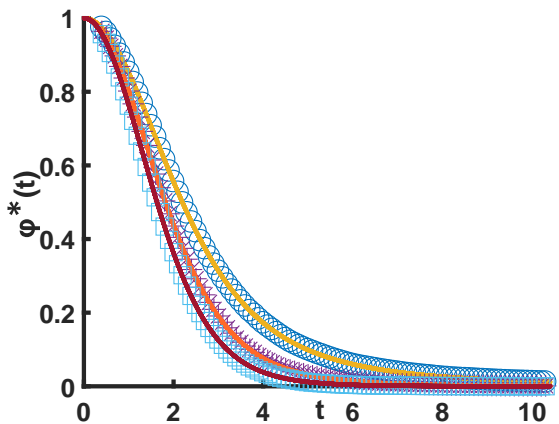}
  \caption{Survival probabilities of return time for the return at a constant acceleration. The PDF of movement time $\psi_m(t)$ is chosen to be power-law distribution and the propagator function $h(x,t)$ is still Gaussian. The acceleration $|a|=1$, the parameter $\tau_0=1$ in $\psi_m(t)$, $\alpha=0.75$ (circles), $\alpha=1.25$ (stars), and $\alpha=1.75$ (squares), respectively. The simulation results are obtained by averaging over $5\times10^4$ realizations.}
  \label{surapowlaw}
\end{figure}

In the following we turn to discuss the case of movement time following power-law distribution $\psi_m=(\alpha/\tau_0) (1+t/\tau_0)^{-1-\alpha}$ with $0<\alpha<1$, and the corresponding survival probability is $\Psi_m(t)=(1+t/\tau_0)^{-\alpha}$. At first, we will focus on the PDF of return time $\varphi(t)$ for each type of return, and the corresponding $n$th moment is also of our interest. For the returning with a constant velocity $v$, from \eqref{1.25} and $h(x,t)$ being a Gaussian distribution, one can obtain 
\cite{Axel}
\begin{equation*}
  \varphi(t)=\frac{v \alpha}{\sqrt{\pi D \tau_0}} \Gamma\bigg(\alpha+\frac{1}{2}\bigg)U\left(\alpha+\frac{1}{2},\frac{1}{2};\frac{v^2 t^2}{4 D \tau_0}\right),
\end{equation*}
where $U(a,b;z)$ is the confluent hypergeometric function of the second kind introduced in Appendix \ref{Appen A}. Besides, from \eqref{mmt_tr} there exists \cite{Axel}
\begin{equation*}
  \langle t_r^n\rangle = \bigg(\frac{\tau_0 D}{v^2}\bigg)^{\frac{n}{2}}\alpha n! U\bigg(\frac{n}{2}+1,\frac{n}{2}+1-\alpha;0\bigg).
\end{equation*}
It can be concluded from \eqref{a2} that $\langle t_r^n\rangle$ is finite only for $\alpha>\frac{n}{2}$. Specifically, the average of return time exists only for $\alpha>1/2$.
%
%
Next for the case of returning in a constant acceleration $|a|$, from \eqref{1.30} we can obtain the PDF of return time
\begin{equation}\label{pdf_cst_a_pw_law}
  \varphi(t)=\frac{\alpha |a| t}{\sqrt{\pi\tau_0 D}}\Gamma\bigg(\alpha+\frac{1}{2}\bigg)U\left(\alpha+\frac{1}{2},\frac{1}{2};\frac{a^2 t^4}{16 D \tau_0}\right);
\end{equation}
and from \eqref{1.31} we have
\begin{equation*}
  \langle t_r^n\rangle=\alpha \left(\frac{4 D \tau_0}{a^2}\right)^{\frac{n}{4}} \Gamma\left(\frac{n}{2}+1\right) U\left(\frac{n}{4}+1,\frac{n}{4}+1-\alpha; 0 \right).
\end{equation*}
Therefore, the $n$th moment of return time is finite in this case only for $\alpha>\frac{n}{4}$, which is weaker than the condition required by the previous case of constant velocity. Moreover, from \eqref{pdf_cst_a_pw_law}, the corresponding survival probability is
\begin{equation*}
	\varphi^*(t)=\frac{\Gamma\left(\alpha+\frac{1}{2}\right)}{\sqrt{\pi}} U\left(\alpha,\frac{1}{2};\frac{a^2 t^4}{16 D \tau_0}\right),
\end{equation*}
which can be verified in Fig. \ref{surapowlaw} through numerical simulations.

\begin{figure}[htbp]
  \centering
  \includegraphics[scale=0.28]{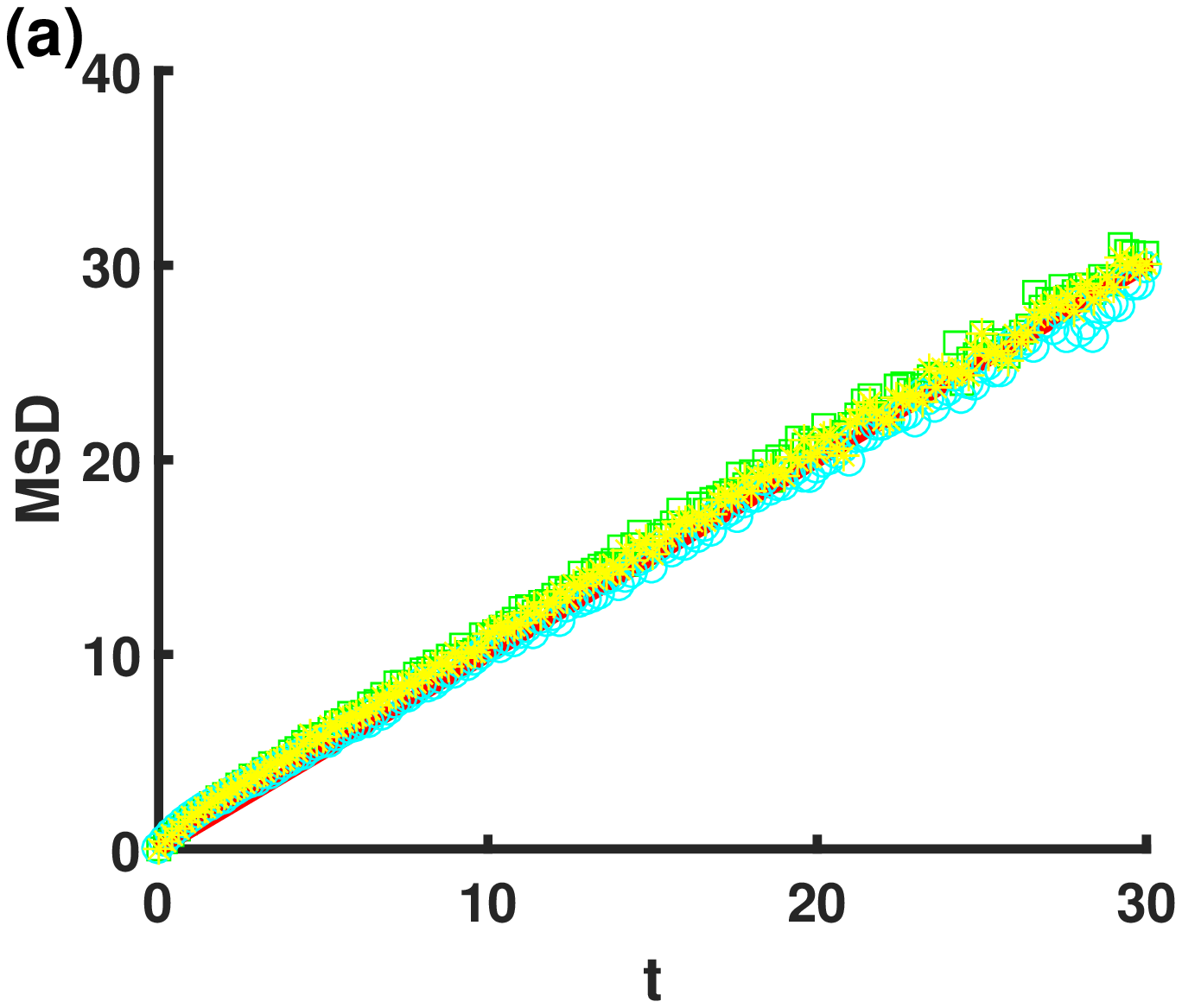}
  \includegraphics[scale=0.28]{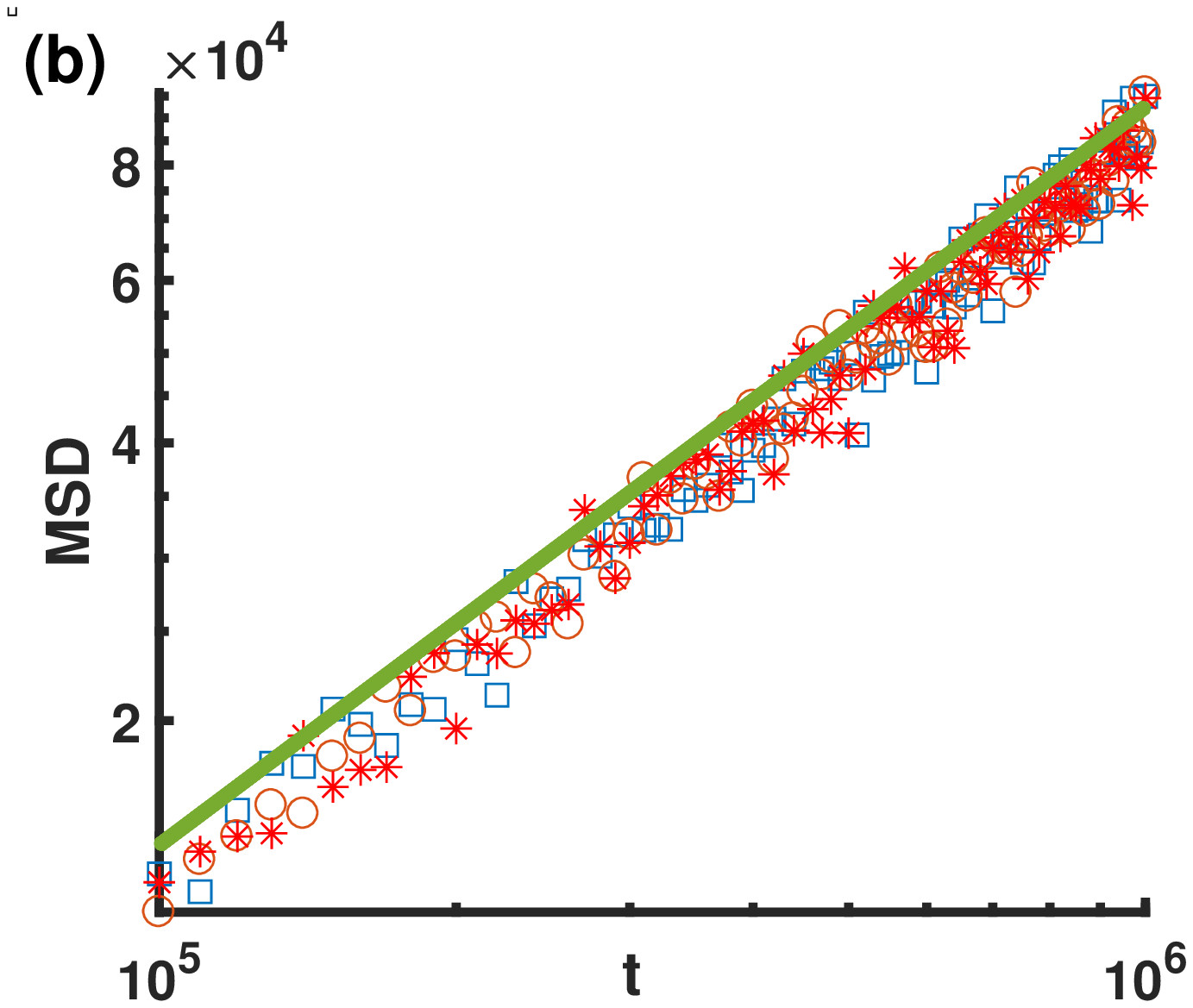}
  \includegraphics[scale=0.28]{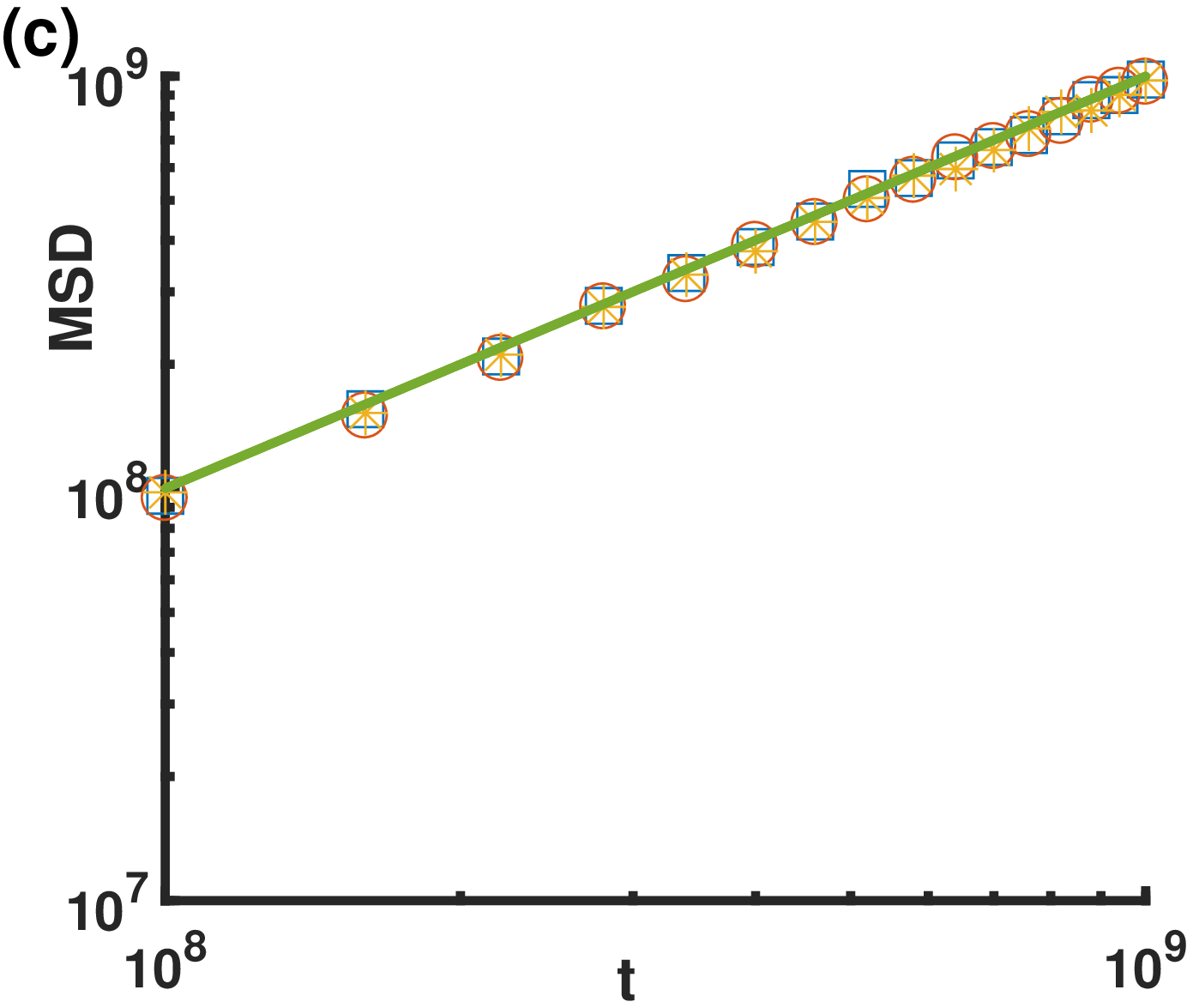}
  \caption{MSDs of the particles returning in three different ways for power-law distributed movement time. The propagator function $h(x,t)$ is still Gaussian and $\psi_m(t)$ is power-law distribution with $\tau_0=1$ and $\alpha=0.5$; other parameters are $v=1$ (squares), $a=1$ (stars), and $k=1$ (circles), respectively. For (a) the rest time is exponentially distributed with $\lambda=1$. For (b) and (c) (in log-log scale), the PDFs of rest time are assumed to be power-law with $\beta=0.3$ and $\beta=0.7$, respectively. The simulation results are obtained by averaging over $10^4$ realizations.}
  \label{msdgaussianpower}
\end{figure}

Next we still consider three different ways of returning to calculate each MSD. For the exponentially distributed rest time $\phi_r(t)=\lambda e^{-\lambda t}$ or constant rest time $\phi_r(t)=\delta(t-t_{\rm rest})$, by utilizing the confluent hypergeometric function of the second kind and combining \eqref{MSDv}, \eqref{MSDa} with \eqref{MSDk}, for the three ways of returning we have
\begin{equation}\label{2.16}
	\langle x^2 (t)\rangle \sim 2D(1-\alpha)t.
\end{equation}
The result of \eqref{2.16} is consistent with \cite{Axel} which has no phase of rest. Therefore, the rest is too slight comparing with the phase of movement in this case.
 Moreover, we consider the PDF $\phi_r(t)$ of rest time being power-law distribution \eqref{pw-law_PDF} with the three different types of returning. The asymptotic behaviors of overall MSDs of processes with the three kinds of returns are the same
\begin{equation}\label{2.17}
\langle x^2 (t)\rangle \sim
\begin{cases}
\frac{ 2D \tau_0^{\alpha-\beta} \Gamma(2-\alpha)}{\Gamma(1-\beta)\Gamma(2-\alpha+\beta)} t^{1-\alpha+\beta}, & \mbox{if $\alpha>\beta$;} \\
  2D(1-\alpha)t,  & \mbox{if $\alpha<\beta$},
\end{cases}
\end{equation}
which is verified by Fig. \ref{msdgaussianpower}.
The asymptotic behavior of MSD in \eqref{2.17} indicates the competition between movement phase and rest phase.

In summary, there is always a competition between the movement phase and rest phase. When the time of movement phase follows power-law distribution with infinite average, then the exponentially distributed or constant rest time or other kind of rest time with finite average can hardly compete with movement time, so that the overall MSD always behaves like \eqref{2.16}. On the other hand, if the average rest time is also divergent, like power-law here we choose, then it can be concluded from \eqref{2.17} that the PDF with heavier tail will control the asymptotic behavior for long time. Moreover, \eqref{2.17} indicates the subdiffusion for $\alpha>\beta$.

%

\section{Movement Phase Following Dynamics of L\'evy Walks}\label{sec 4}

L\'{e}vy walk \cite{First,Zaburdaev} is one of the most popular stochastic processes to model anomalous diffusion, especially super-diffusion. A representative example of L\'evy walks is the one with a finite velocity $v_0$ for each step of movement, so that the length of each step of movement is $v_0 \tau$ with $\tau$ being the duration of that step. The durations of each step of movement before last step of exceeding observation time $t$ are i.i.d., the corresponding PDF of which is denoted as $\psi(\tau)$. The L\'evy walks with instant stochastic resetting have been studied in \cite{Tian}, which can be considered an instant return to the origin or other fixed position. Moreover, L\'{e}vy walks under the action of external harmonic forces have been analyzed in \cite{Bo} by utilizing Hermite polynomials. In the following we consider the process with movement, return, and rest phases, where the movement phase follows the dynamics of symmetric one dimensional L\'evy walks with constant velocity $v_0$.
%

As shown in \cite{Zaburdaev}, with the help of Fourier transform defined as
\begin{equation*}
	\tilde{f}(\rho)=\mathscr{F}_{x\to \rho}\{f(x)\}=\int_{-\infty}^{\infty} e^{- i \rho x} f(x) dx,
\end{equation*}
the PDF $P_l(x,t)$ of L\'{e}vy walk considered here can be explicitly given in the Fourier-Laplace space
\begin{equation}\label{PDF_LW}
\begin{split}
\widehat{\widetilde{P}}_l(\rho, s) &= \mathscr{F}_{x\to \rho}\left\{\mathscr{L}_{t\to s}\left\{P_l(x,t) \right\} \right\} \\
& = \frac{\left[\widehat{\Psi}(s+i k v_0)+\widehat{\Psi}(s-i k v_0)\right] \widetilde{P}_0(k)}{2-\left[\hat{\psi}(s+i k v_0)+\hat{\psi}(s-i k v_0)\right]},
\end{split}
\end{equation}
where $\widetilde{P}_0(k)$ represents the Fourier transform of the initial condition $P_0(x)$; specially we choose $P_0(x)=\delta(x)$ so that $\widetilde{P}_0(k)=1$. And $\Psi(t)$ in the above equation defined as $\Psi(t) = \int_{t}^{\infty} \psi(\tau) d\tau$ represents the survival probability.

\begin{figure}[htbp]
  \centering
  \includegraphics[width=8cm]{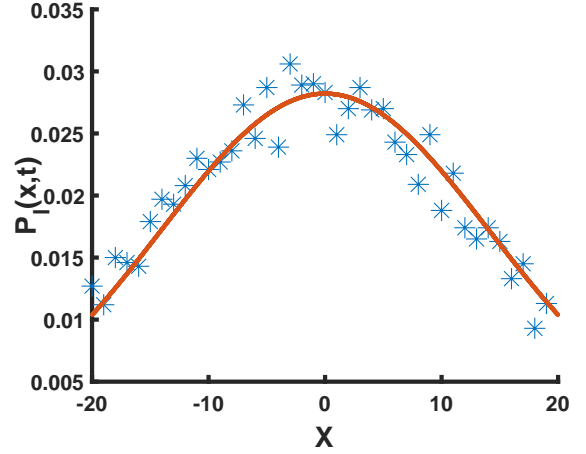}
  \caption{PDF of L\'{e}vy walk with exponentially distributed walking time. The parameters are $v_0=1$, $x_0=0$, and $r=1$ for $\psi(\tau)$. The simulation results are obtained by averaging over $10^4$ realizations.}
  \label{pdfoflevywalk}
\end{figure}

When the walking time is exponentially distributed $\psi(\tau) = r e^{-r \tau}$ with $r>0$, then the Fourier-Laplace transform of PDF $P_l(x,t)$ can be simplified from \eqref{PDF_LW} as
\begin{equation*}
  \widehat{\widetilde{P}}_l(k,s)=\frac{r+s}{s(r+s)+k^2v_0^2}.
\end{equation*}
Further after taking inverse Fourier transform and utilizing the fact $\mathscr{F}_{x\rightarrow k}\{e^{-a |x|}\}=\frac{2 a}{a^2+k^2}$, we have
\begin{equation*}
	\widehat{P}_l(x,s)=\frac{1}{2 v_0}\sqrt{\frac{r+s}{s}} \exp\left(-\sqrt{\frac{s (r+s)}{v_0^2}}|x| \right).
\end{equation*}
Therefore, for sufficiently large time $t$ (indicating small $s$ in frequency domain), we have the asymptotic behavior of PDF ${P}_l(x,t)$ in Laplace space  $\widehat{P}_l(x,s) \sim \frac{1}{2 v_0}\sqrt{\frac{r}{s}} \exp\left( -\sqrt{\frac{s r}{v_0^2}}|x| \right)$, which leads to
\begin{equation}\label{3.4}
  P_l(x,t) \sim \frac{\sqrt{r}}{2 v_0 \sqrt{\pi t}}\exp\left(-\frac{r x^2}{4 v_0^2 t} \right).
\end{equation}
The asymptotic form of \eqref{3.4} can also be considered as a Gaussian distribution; some statistical properties such as MSD of the process with three phases can be calculated by the methods in Section \ref{sec 3}. The result of \eqref{3.4} is numerically verified in Fig. \ref{pdfoflevywalk}.

\begin{figure}[htbp]
  \centering
  \includegraphics[scale=0.28]{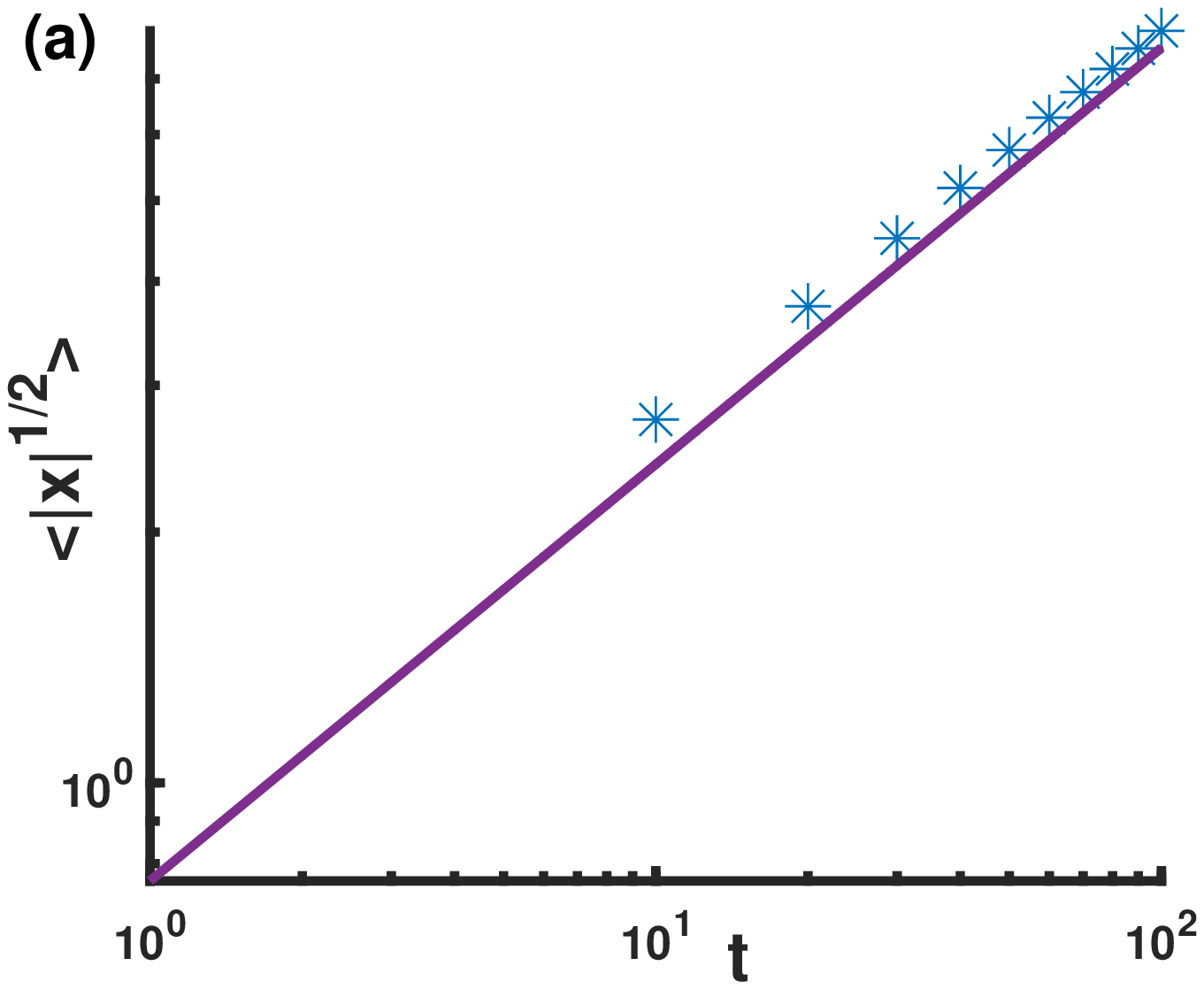}
  \includegraphics[scale=0.28]{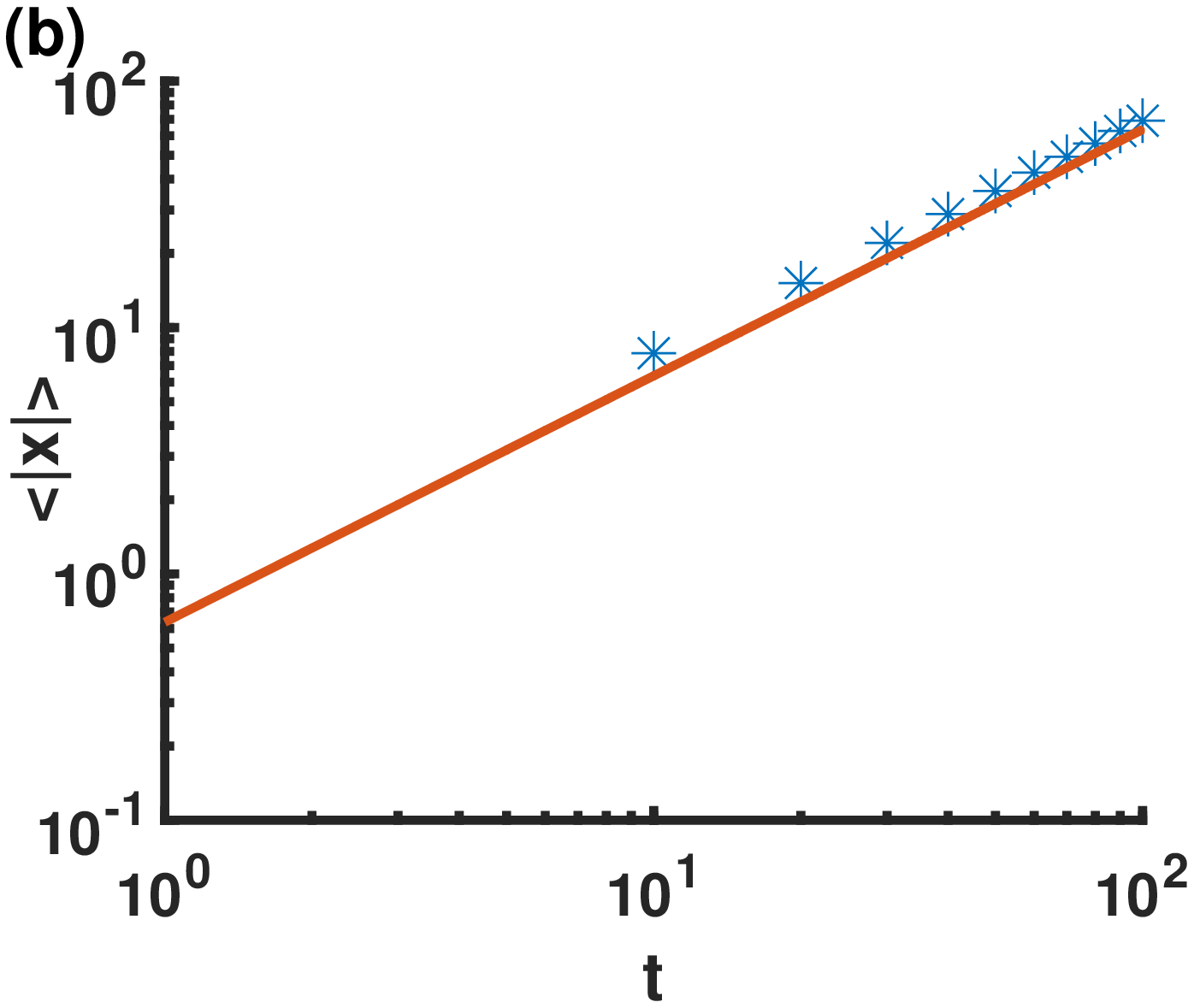}
  \includegraphics[scale=0.28]{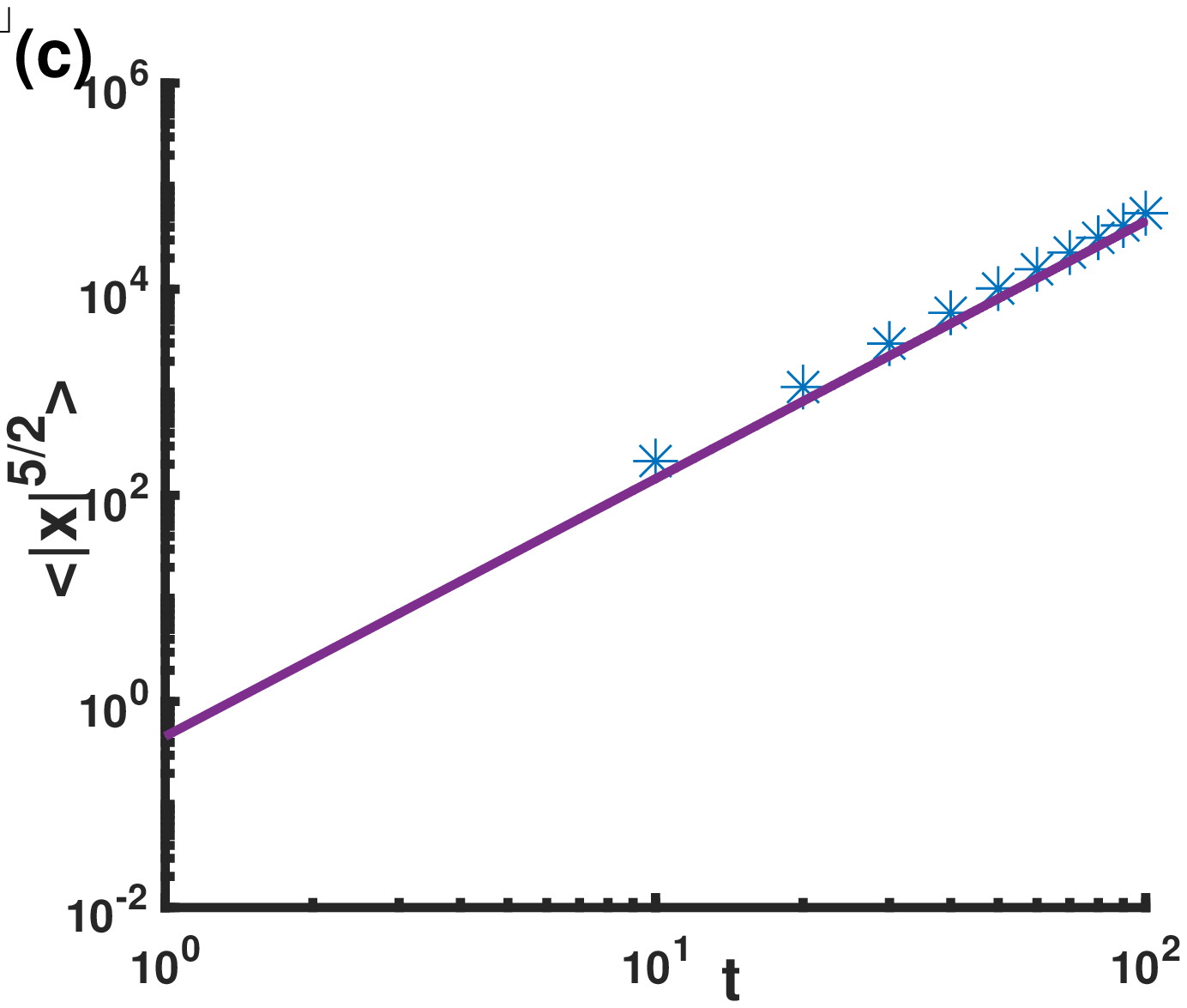}
  \includegraphics[scale=0.28]{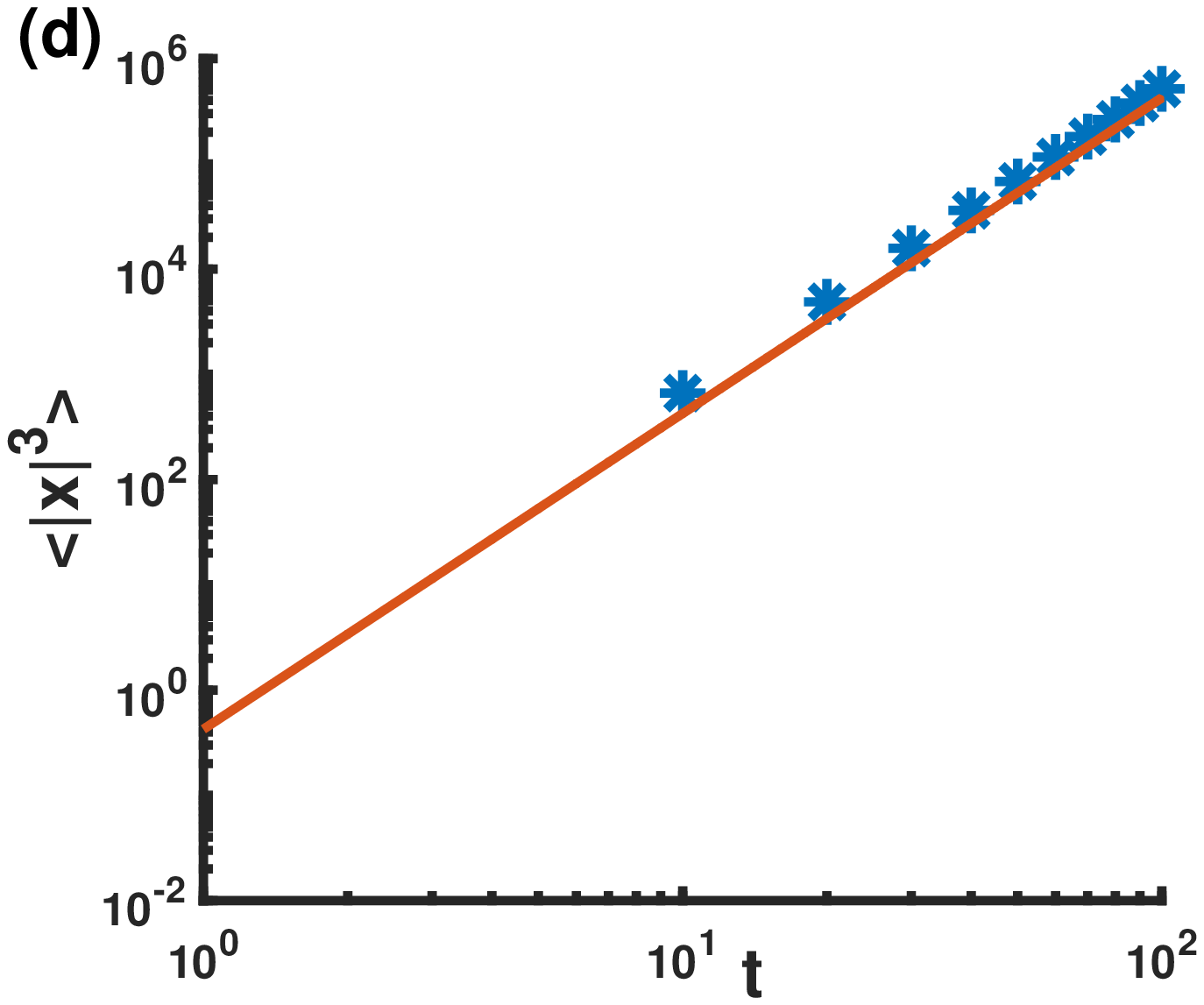}
  \caption{Numerical simulations of the $n$th and $(n/2)$th moments of absolute value of position of L\'{e}vy walk with power-law walking time $\gamma=\frac{1}{2}$ (in log-log scale). The parameters are $v_0=\tau_0=1$, $x_0=0$. The real lines in (a) and (c) are the theoretical results in \eqref{b2}, while real lines in (b) and (d) are obtained by \eqref{b1}. The simulation results marked in stars are obtained by averaging over $10^4$ realizations.}
  \label{moment}
\end{figure}

Next we consider the L\'evy walk with walking time PDF $\psi(\tau) = \gamma/\tau_0 (1+\tau/\tau_0)^{-1-\gamma}$, $0<\gamma<1$. Then from \cite{Froemberg}, the asymptotic form of PDF in time domain can be explicitly given by
\begin{equation*}
\begin{split}
    &P_l(x,t) \sim \frac{\sin(\pi \gamma)}{\pi}\times\\
     & \frac{(v_0 t-x)^\gamma(v_0 t+x)^{\gamma-1}\!+\!(v_0 t+x)^\gamma(v_0 t-x)^{\gamma-1}}{(v_0 t+x)^{2\gamma}\!+\!(v_0 t-x)^{2\gamma}\!+\!2 (v_0 t+x)^{\gamma}(v_0 t-x)^{\gamma}\cos(\pi \gamma)},
\end{split}
\end{equation*}
for $|x| \leq v_0 t$. 
For the special case $\gamma=\frac{1}{2}$, the corresponding PDF can be further simplified as
\begin{equation}\label{3.7}
  P_l(x,t) \sim \frac{1}{\pi \sqrt{v_0^2t^2-x^2}}.
\end{equation}
From \eqref{3.7}, the $n$th and $(n/2)$th moments of absolute values of positions of L\'{e}vy walk can be explicitly obtained as
\begin{equation}\label{b1}
  \langle |x|^n(t) \rangle \sim \frac{2 v_0^n \Gamma(\frac{1+n}{2})}{n \sqrt{\pi} \Gamma(\frac{n}{2})}t^n
\end{equation}
and
\begin{equation}\label{b2}
  \langle |x|^{n/2}(t)\rangle \sim \frac{v_0^{\frac{n}{2}} \Gamma(\frac{2+n}{4})}{\sqrt{\pi} \Gamma(1+\frac{n}{4})} t^{\frac{n}{2}}.
\end{equation}
The results of \eqref{b1} and \eqref{b2} are consistent with the numerical simulations in Fig. \ref{moment}.

%
%
%

%

\subsection{Exponentially distributed movement time}

\begin{figure}[htbp]
  \centering
  \includegraphics[scale=0.28]{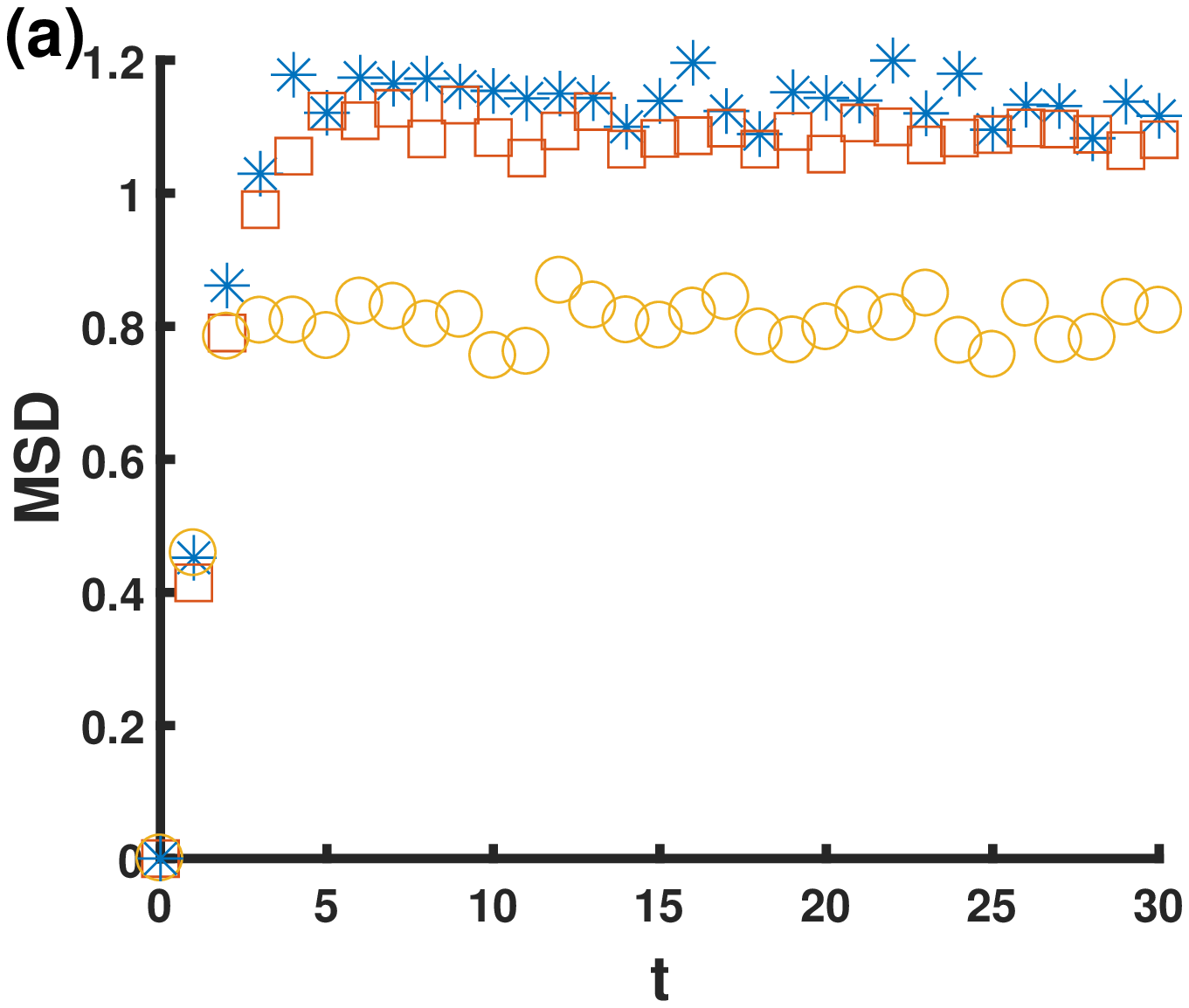}
  \includegraphics[scale=0.28]{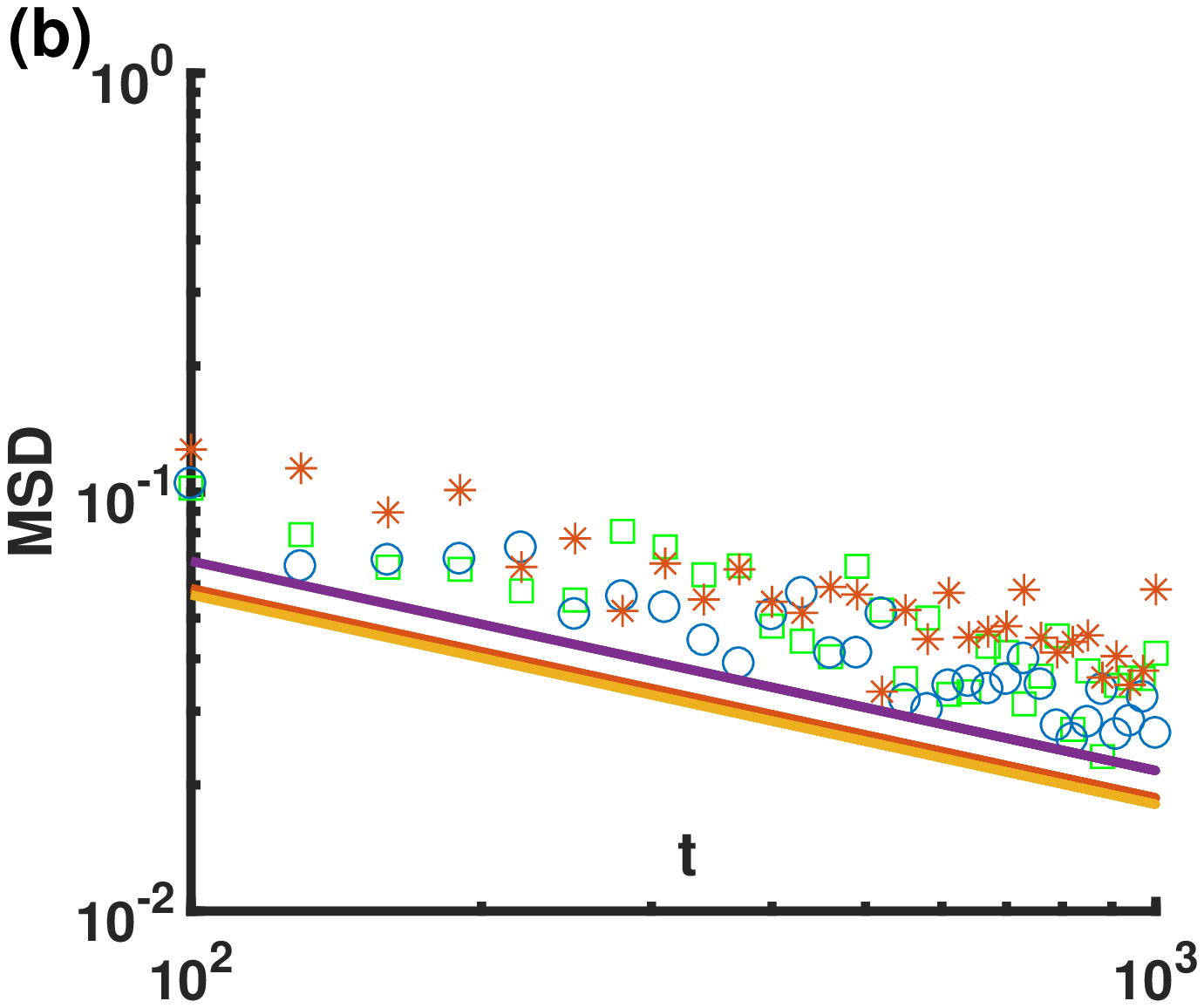}
  \caption{MSDs for the processes with L\'evy walk kind of propagator functions and exponentially distributed movement time. Here $\gamma=1/2$ in PDF $\phi(\tau)$ of walking duration of each step of L\'evy walk process. The PDF of movement time $\psi_m(t)=e^{-t}$. The other parameters are $v=1$ (squares), $a=1$ (stars), and $k=1$ (circles), respectively. For (a) the PDF of rest time $\phi_r(t)$ is also exponential distribution with $\lambda=1$. For (b) (in log-log scale), $\phi_r(t)$ is assumed to be power-law distribution with $\beta=0.5$, which turns out to be the same slope between the theoretical results (real lines) and numerical simulations. The simulation results are obtained by averaging over $10^4$ realizations. }
  \label{msdlevyexp}
\end{figure}

We consider that the movement time follows exponential distribution, i.e., $\psi_m(t)=\alpha e^{-\alpha t}$, while the PDF of rest time is $\phi_r(t)=\lambda e^{-\lambda t}$. Besides, let the propagator function $h(x,t)=P_l(x,t)$ in \eqref{3.7}, so that $\langle |x|^n(t) \rangle_h$ and $\langle |x|^{n/2}(t)\rangle_h$ are given in \eqref{b1} and \eqref{b2}, respectively. Similar to the discussions in Section \ref{sec 3}, the overall MSDs for the processes returning in a constant velocity, acceleration, and under a harmonic force can be obtained from \eqref{MSDv}, \eqref{MSDa}, and \eqref{MSDk}, respectively. According to \eqref{b1} and \eqref{b2} we can get the asymptotic behaviors of overall MSDs for different types of returning when time $t$ is large enough, and it turns out that all MSDs in this case have the same asymptotic behavior $\langle x^2(t)\rangle \sim t^0$, i.e., the processes are localized, which are verified by the numerical simulations (see Fig. \ref{msdlevyexp}(a)). However, it is not easy to give the accurate constants of MSDs in this case.



We further take PDF $\phi_r(t)$ of rest time to be power-law in \eqref{pw-law_PDF}. Similarly, the asymptotic behavior of the overall MSD for returning at constant velocity $v$ (please note that $v_0$ represents constant velocity for L\'evy walk process in movement phase) is
\begin{equation*}
  \langle x^2(t)\rangle \sim \frac{v_0^2}{\alpha^3}\frac{\sin(\pi\beta)}{\pi \tau_0^\beta}\bigg(1+\frac{8 v_0}{3 v \pi}\bigg)t^{\beta-1}.
\end{equation*}
For returning at constant acceleration $|a|$, the corresponding asymptotic behavior of MSD is
\begin{equation*}
  \langle x^2(t)\rangle\sim
   \frac{v_0^2}{\alpha^3}\frac{\sin(\pi\beta)}{\pi \tau_0^\beta}
   \left(1+\frac{12\Gamma\left(\frac{3}{4}\right) }{5\Gamma\left(\frac{1}{4}\right)}\sqrt{\frac{2 v_0 \alpha}{|a| }}\right)t^{\beta-1}.
\end{equation*}
As for the case of returning under a harmonic force, there exists
\begin{equation*}
  \langle x^2(t)\rangle\sim
   \frac{v_0^2}{\alpha^3}\frac{\sin(\pi\beta)}{\pi \tau_0^\beta}\left(1+\frac{\pi \alpha}{4 \sqrt{k}}\right)t^{\beta-1}.
\end{equation*}
The above results are verified in Fig. \ref{msdlevyexp}(b).

\subsection{Power-law distributed movement time}

\begin{figure}[htbp]
  \centering
  \includegraphics[scale=0.28]{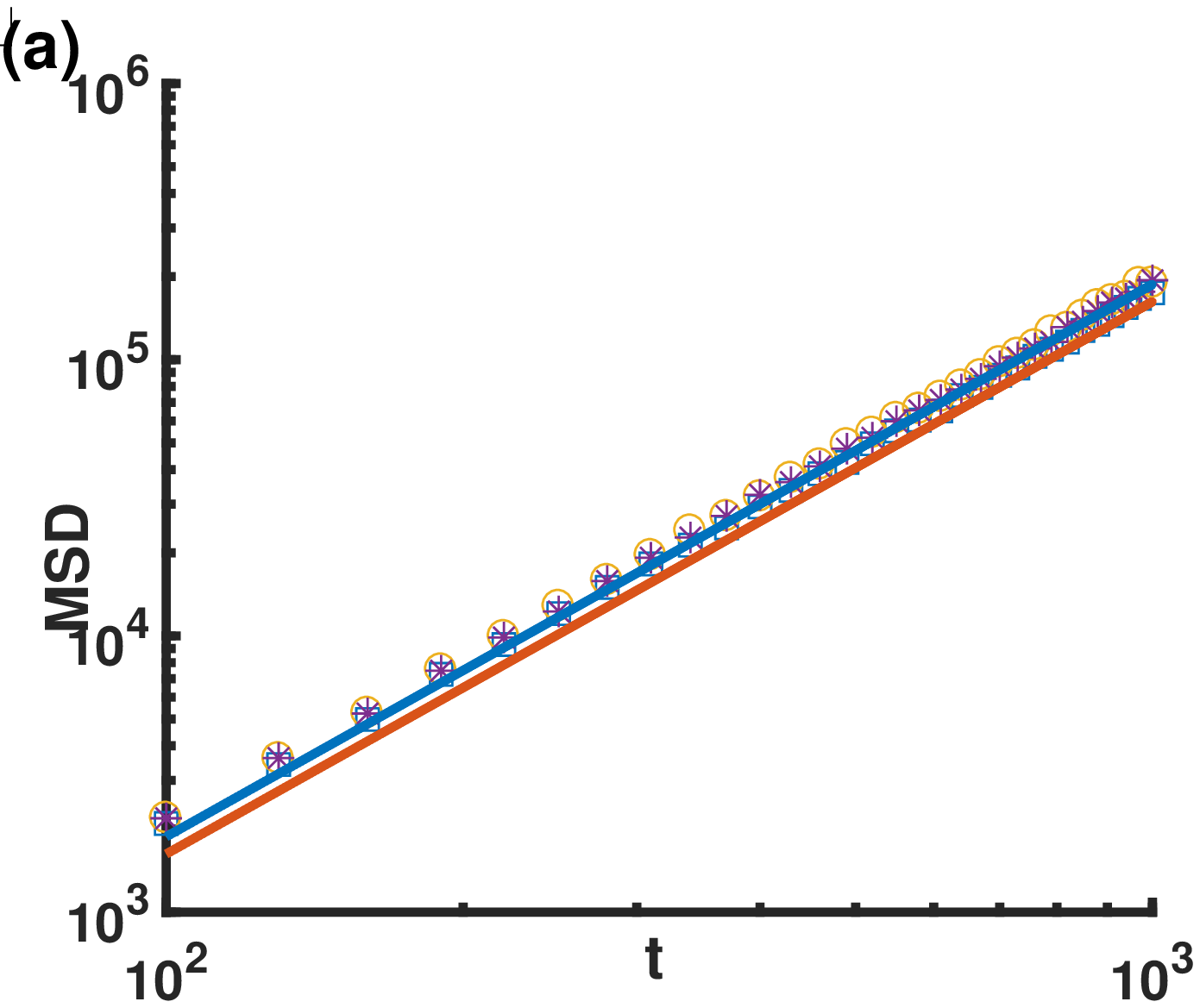}
  \includegraphics[scale=0.28]{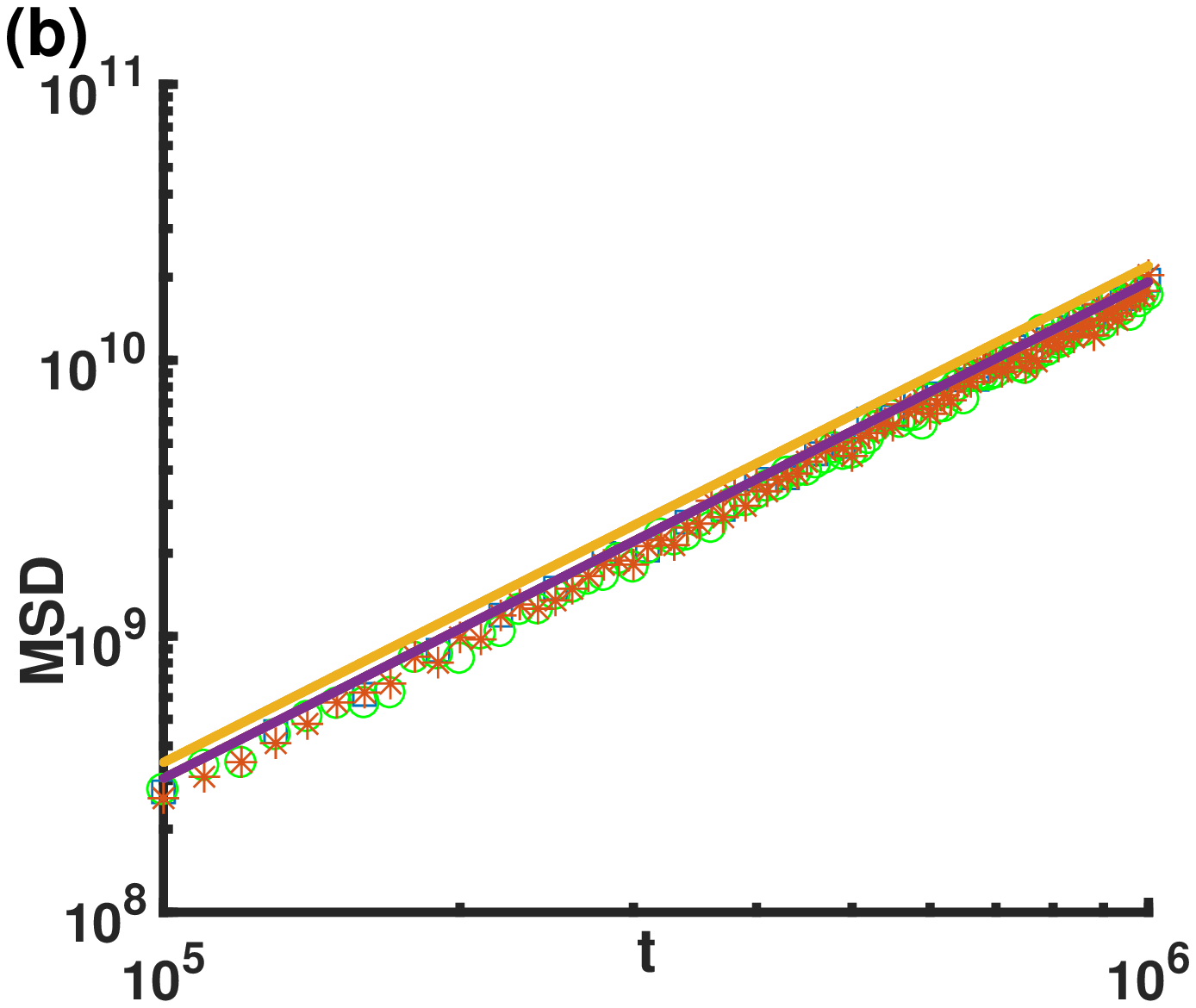}
  \includegraphics[scale=0.28]{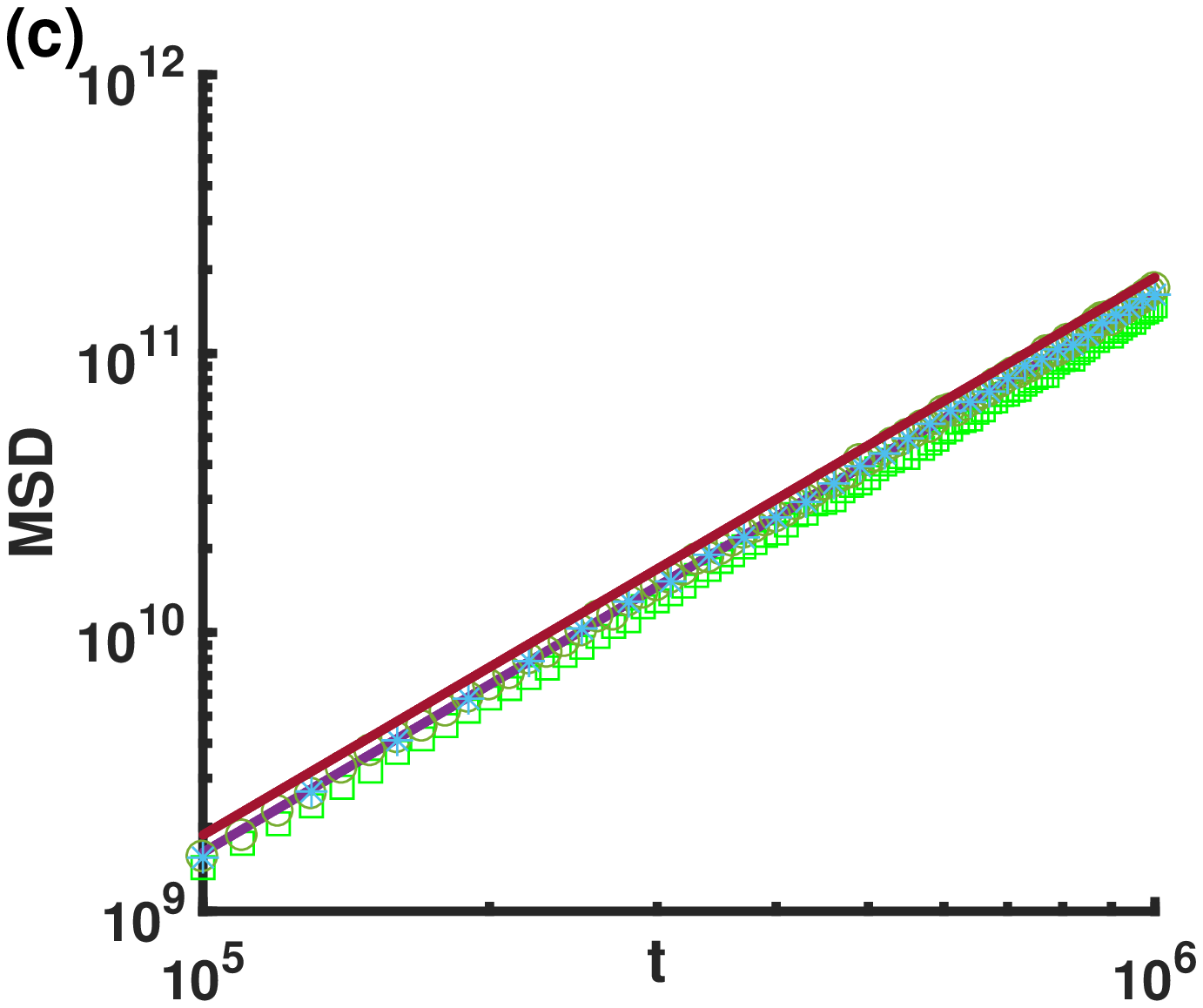}
  \caption{MSDs for the processes with L\'evy walk kind of propagator functions and power-law distributed movement time (in log-log scale). The walking duration for L\'evy walk in movement phase still follows power-law distribution with $\gamma=1/2$, and the PDF of reset time is assumed to be power-law shown in \eqref{pw-law_PDF} with $\alpha=0.5$; the other parameters are $v=1$ (squares), $a=1$ (stars), and $k=1$ (circles), respectively. For (a) the PDF of rest time $\phi_r(t)$ is exponential distribution with $\lambda=1$. For (b) and (c) $\phi_r(t)$ is assumed to be power-law distribution, and $\beta=0.3$ and $0.7$, respectively. The simulation results are obtained by averaging over $10^4$ realizations.}
  \label{msdlevypowlaw}
\end{figure}

In this part, we mainly consider the particle whose movement time follows power-law distribution, $\psi_m(t)=\frac{1}{\tau_0}\frac{\alpha}{(1+t/\tau_0)^{1+\alpha}}$, $0<\alpha<1$. For the case of returning at a constant velocity $v$ and exponentially distributed rest time $\phi_r(t)=\lambda e^{-\lambda t}$, from \eqref{MSDv} we have
\begin{equation}\label{MSD_LW_PW_V_exp}
   \langle x^2(t)\rangle \sim \frac{(2-\alpha)(1-\alpha)v_0^2}{2}\frac{9 v \pi+8\alpha v_0}{18 v \pi+36 \alpha v_0}t^2.
 \end{equation}
 Moreover, if the PDF of rest time $\phi_r(t)$ is power-law distribution in \eqref{pw-law_PDF}, then we have
\begin{equation}\label{MSD_LW_PW_V_PW}
\langle x^2(t)\rangle\!\sim
  \begin{cases}
   \frac{\Gamma(3-\alpha) \tau_0^{\alpha-\beta}}{\Gamma(3+\beta-\alpha)\Gamma(1-\beta)}\!\left(\frac{v_0^2}{2}\!+\!\frac{4 \alpha v_0^3}{9 v \pi}\!\right)t^{2+\beta-\alpha}, & \mbox{if $\alpha\!>\!\beta$};\\
    \frac{(2-\alpha)(1-\alpha)v_0^2}{2}\frac{9 v \pi+8\alpha v_0}{18 v \pi+36 \alpha v_0}t^2, & \mbox{if $\alpha\!<\!\beta$}.
  \end{cases}
\end{equation}
The results of \eqref{MSD_LW_PW_V_exp} and \eqref{MSD_LW_PW_V_PW} are verified in Fig. \ref{msdlevypowlaw}. Besides, the overall MSDs have the same asymptotic behaviors when the particles return at a constant acceleration and under the action of harmonic force. Specifically, substituting exponential distribution of $\phi_r(t)$ with average $1/\lambda$ into \eqref{MSDa} and \eqref{MSDv}, we can obtain the same asymptotic result
\begin{equation}\label{3.13}
  \langle x^2(t)\rangle\sim
    \frac{v_0^2}{4}(2-\alpha)(1-\alpha)t^2.
\end{equation}
While for the power-law distributed rest time with PDF $\phi_r(t)$ shown in \eqref{pw-law_PDF},
we have
\begin{equation}\label{3.14}
  \langle x^2(t)\rangle\sim
  \begin{cases}
  \frac{v_0^2}{2}\frac{\Gamma(3-\alpha) \tau_0^{\alpha-\beta}}{\Gamma(3+\beta-\alpha)\Gamma(1-\beta)}t^{2+\beta-\alpha}, & \mbox{if $\alpha>\beta$ };\\
    \frac{v_0^2}{4}(2-\alpha)(1-\alpha)t^2, & \mbox{if $\alpha<\beta$ } .
  \end{cases}
\end{equation}

In summary, for power-law distributed movement time, if the process returns at a constant velocity $v$, then the asymptotic behavior of overall MSD will be influenced by $v$. Whereas, if returning at a constant acceleration $|a|$ or under action of harmonic force, the overall MSDs have the same asymptotic form, and $|a|$ and parameter $k$ of harmonic force have no influence on the MSD after sufficiently long time. From \eqref{MSD_LW_PW_V_PW} and \eqref{3.14}, the super diffusions can be observed for $\alpha>\beta$ while ballistic diffusions appear for $\alpha<\beta$.
%
%

\section{Mean First passage time}\label{sec 5}

In the above, we have discussed the asymptotic behaviors of the process with three different phases. In this section, we consider another important quantity for stochastic process, which is mean first passage time (MFPT) \cite{KGM2014,DWW2017}. First passage time (FPT) is defined as the first time arriving at the boundary of a given domain. Here we consider the domain as $(-\infty, b]$ with $b>0$, so the first time when the process reaches position $x=b$ is the FPT we consider here, denoted as $t_f$. Obviously, the event of reaching $x=b$ only happens during the movement phase.

The PDF $q(t)$ of FPT can be obtained from the survival probability $G_b(t)$ of overall PDF $P(x,t)$ on the domain $(-\infty, b]$, which is $G_b(t)=\int_{-\infty}^b P(x,t)dx$. Then the cumulative distribution function $Q(t)$ of FPT has the following relation with $G_b(t)$ for $x=b$ being the absorbing boundary \cite{G2004,DS2015},
\begin{equation*}
	Q(t)=\int_0^t q(u)du=1-G_b(t),
\end{equation*}
which indicates $q(t)=-\frac{d}{d t}G_b(t)$. Therefore, for the initial condition $P(x,t=0)=\delta(x)$, in Laplace space we have $\hat{q}(s)=-s \widehat{G}_b(s)+1$. Finally, the average of FPT can be expressed by
\begin{equation}\label{MFPT_Gb}
	\begin{split}
		\langle t_f\rangle = -\lim_{s \to 0}\frac{d}{ds}\hat{q}(s)=\lim_{s \to 0} \widehat{G}_b(s).
	\end{split}
\end{equation}
In this section, we consider the case that the dynamic of movement phase is Brownian motion. Since $x=b$ is an absorbing boundary, then the propagator function $h_b(x,t)$ satisfies
\begin{equation*}
	\left\{
	\begin{split}
		&\frac{\partial}{\partial t} h_b(x,t) = \frac{\partial^2}{\partial x^2}h_b(x,t),\quad \text{for $x<b$;}\\
		&h_b(x=b,t) = 0.
	\end{split}\right.
\end{equation*}
Then, by solving the above equation, we have
\begin{equation}\label{4.8}
  h_b(x,t)=\frac{1}{\sqrt{4 \pi D t}}\left[e^{-\frac{x^2}{4 D t}}-e^{-\frac{(x-2 b)^2}{4 D t}}\right] \quad \text{for $x\leq b$.}
\end{equation}

%

According to dynamics of the process with three phases, the survival probability $G_b(t)$ can be shown as
\begin{widetext}
\begin{equation}\label{4.2}
\begin{split}
   G_b(t) = &\Psi_m(t) \int_{-\infty}^{b} h_b(x,t) dx + \int_{0}^{t} \psi_m (u) du \int_{-\infty}^{b}h_b(x,u) dx \int_{t-u}^{\infty} \delta\left(\theta - R^{-1}(0;x)\right)d\theta\\
     & +\int_{0}^{t} \psi_m(u) du \int_{-\infty}^{b} h_b(x,u) dx \int_{0}^{t-u} \delta\left(\theta-R^{-1}(0;x)\right)\Phi_r(t-u-\theta)d\theta \\
     & +\int_{0}^{t} \psi_m(u) du \int_{-\infty}^{b} h_b(x,u) dx\int_{0}^{t - u} \delta\left(\theta - R^{-1}(0;x) \right) d\theta \int_{0}^{t-u-\theta} \phi_r(\eta) G_b(t - u - \theta - \eta) d\eta.
\end{split}
\end{equation}
\end{widetext}
The r.h.s. of \eqref{4.2} can be interpreted as follows. The first term represents the probability that the particle keeps in movement phase during the whole time to $t$ and moves in the domain $(-\infty,b]$. The second term is for the probability that the particle finishes its movement before time $t$ and arrives at $x<b$ and in the left time keeps returning to the origin. The third term means the probability that the particle has not escaped from semi-infinite domain and stays in rest phase at time $t$. The last term represents that there are at lease one renewal processes during the whole observation time $t$.
%
%
Denote $H_b(t)=\int_{-\infty}^b h_b(x,t)dx$ and
\begin{equation*}
  \varphi_b(t) = \int_{-\infty}^{b}\delta\left(t-R^{-1}(0;x)\right)dx \int_{0}^{\infty}\psi_m(u) h_b(x,u) du,
\end{equation*}
representing PDF of return time for movement phase following dynamics of Brownian motion in $(-\infty, b]$ and absorbing at $x=b$. Then the Laplace transform of $\varphi_b(t)$ is
\begin{equation*}
	 \hat{\varphi}_b(s) = \int_{0}^{\infty} \psi_m(u) du \int_{-\infty}^{b} e^{-s R^{-1}(0;x)}h_b(x,u) dx.
\end{equation*}
Following the derivations in Appendix \ref{app_b}, the Laplace transform of \eqref{4.2} yields
\begin{equation}\label{surhatGb}
\begin{split}
   \widehat{G}_b(s)  &\sim \frac{\mathscr{L}_{t\rightarrow s}\left\{\Psi_m(t) H_b(t)\right\}-\hat{\varphi}_b'(s)|_{s=0}}{1-\hat{\phi}_r(s)\mathscr{L}_{t\rightarrow s}\{\psi_m(t)H_b(t)\}} \\
    &+\frac{\widehat{\Phi}_r(s)\mathscr{L}_{t\rightarrow s}\{\psi_m(t)H_b(t)\}}{1-\hat{\phi}_r(s)\mathscr{L}_{t\rightarrow s}\{\psi_m(t) H_b(t)\}}.
\end{split}
\end{equation}
%
From \eqref{4.8}, it can be obtained  that
\begin{equation*}
  H_b (t)={\rm erf}\left(\frac{b}{2\sqrt{D t}}\right),
\end{equation*}
where ${\rm erf}(x)$ is the error function.

\begin{figure}[htbp]
  \centering
  \includegraphics[scale=0.28]{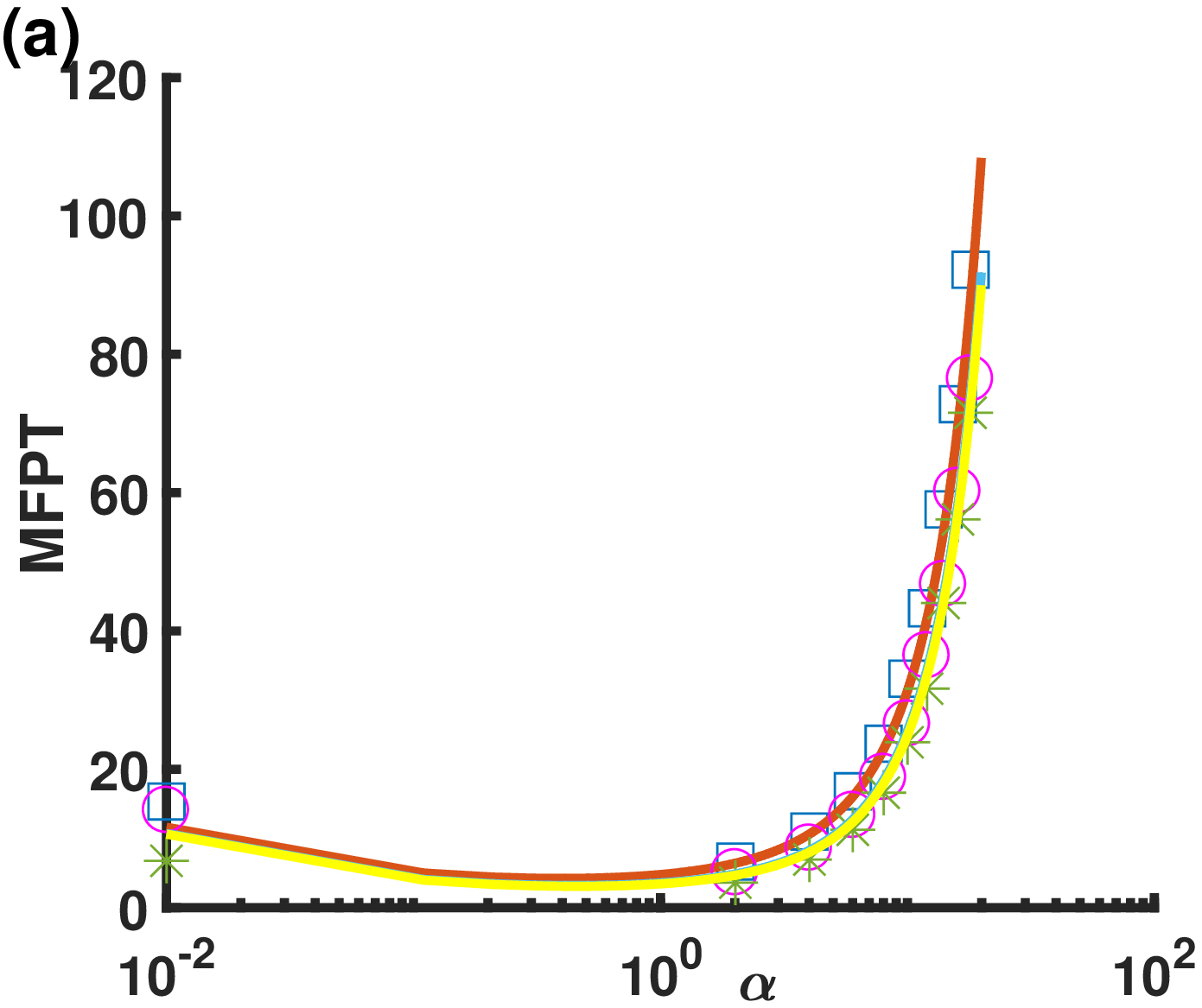}
  \includegraphics[scale=0.28]{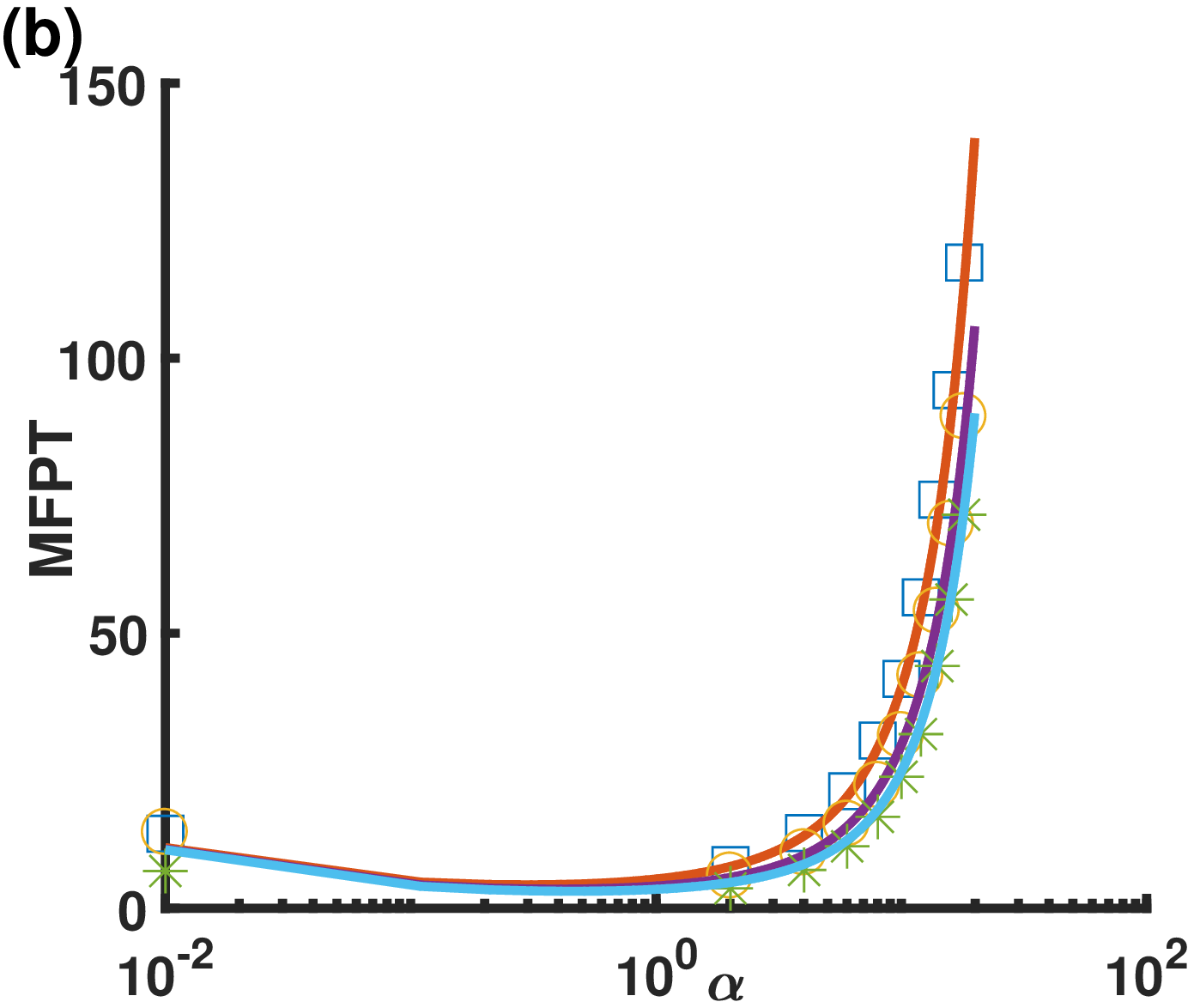}
  \includegraphics[scale=0.28]{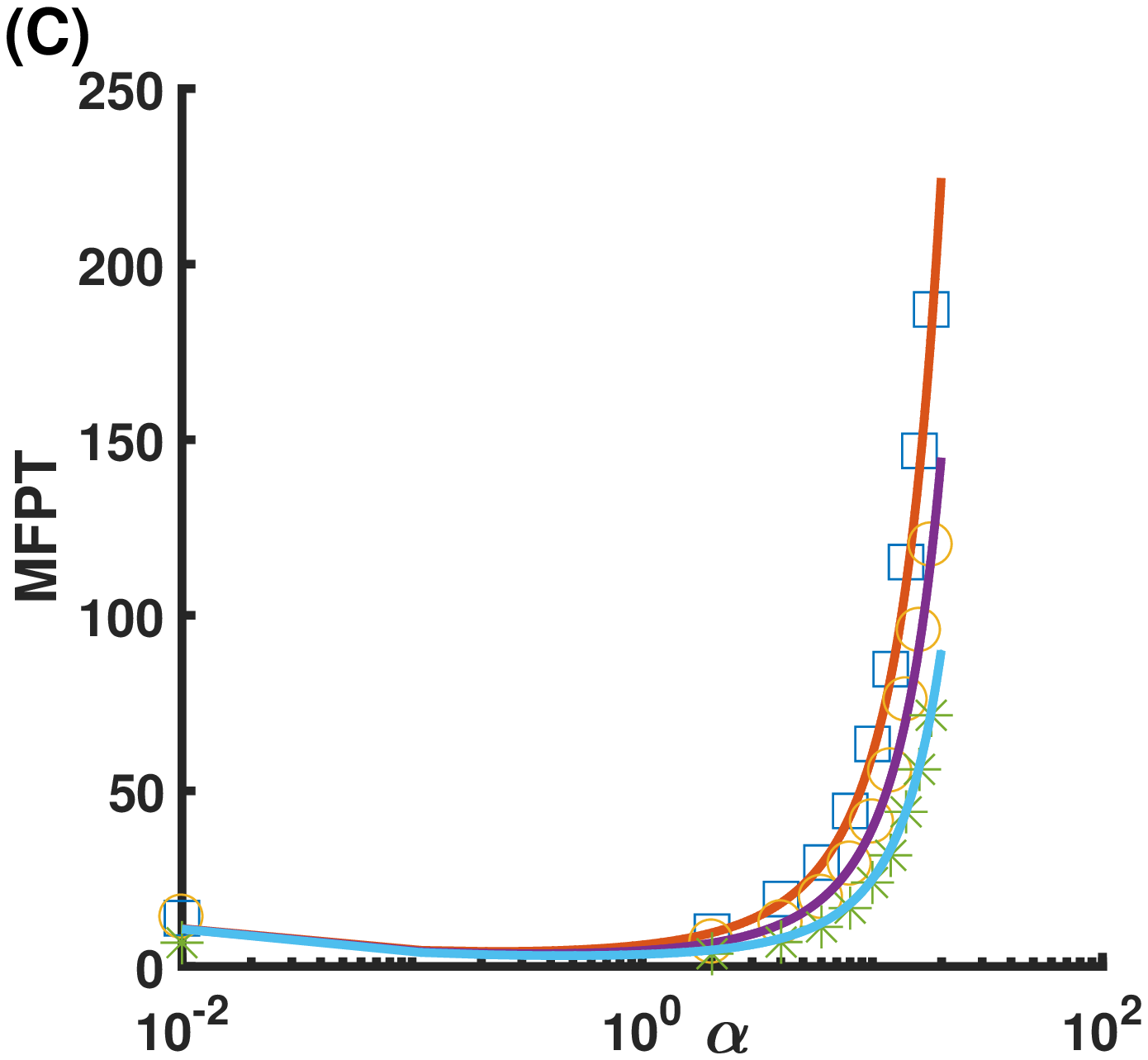}
  \caption{MFPT for the particle returning in three different ways. The domain is $(-\infty,1]$, and $D=1$. Both PDFs of movement time and rest time are chosen to be exponential distribution with unit average. In (a) we choose return at constant velocity $v=1$ (squares) and $v=10$ (circles), respectively. For (b), we simulate particles returning at  constant acceleration  $|a|=1$ (squares) and $|a|=10$ (circles), respectively. And (c) means that the particles return under the action of harmonic force with $k=1$ (squares) and $k=6$ (circles), respectively. While stars in (a) (b) (c) correspond to the case of instantaneous return. The simulation results are obtained by averaging over $10^4$ realizations.}
  \label{MFPT}
\end{figure}

Next we calculate the MFPT $\langle t_f\rangle$ for some specific PDFs of movement time $\psi_m(t)$ and PDFs of rest time $\phi_r(t)$ combining with different kinds of return. First we take $\psi_m(t) = \alpha e^{-\alpha t}$. If $\phi_r(t)$ is chosen to be power-law \eqref{pw-law_PDF}, $\langle t_f\rangle$ is then divergent for all three ways of return considered in this paper. While for the exponentially distributed rest time $\phi_r(t)=\lambda e^{-\lambda t}$ and returning at a constant velocity $v$, the corresponding MFPT can be obtained from \eqref{MFPT_Gb} and \eqref{surhatGb} as
\begin{equation*}
\begin{split}
   \langle t_f \rangle=&\frac{1}{\alpha}\Big(e^{\sqrt{\frac{\alpha}{D}}b}-1\Big)+\sqrt{\frac{D}{\alpha}}\frac{1}{v}\Big(e^{\sqrt{\frac{\alpha}{D}}b}-e^{-\sqrt{\frac{\alpha}{D}} b}\Big) \\
     & -\frac{b}{v}+\frac{1}{\lambda}\Big(e^{\sqrt{\frac{\alpha}{D}}b}-1\Big),
\end{split}
\end{equation*}
which has been verified by numerical simulation in Fig. \ref{MFPT}. Next for the case of returning at constant acceleration $|a|$, there exists
\begin{equation*}
  \begin{split}
     \langle t_f \rangle &= \frac{1}{\alpha}\Big(e^{\sqrt{\frac{\alpha}{D}} b}-1\Big)+ \frac{D^{\frac{1}{4}}\sqrt{\pi}}{2 \sqrt{2} \alpha^{\frac{1}{4}}\sqrt{|a|}}\Big(e^{\sqrt{\frac{\alpha}{D}}b}-e^{-\sqrt{\frac{\alpha}{D}}b}\Big) \\
       &+\frac{1}{\lambda}\Big(e^{\sqrt{\frac{\alpha}{D}} b}-1\Big)+\frac{D^{\frac{1}{4}}\sqrt{\pi}}{2 \sqrt{2}\alpha^{\frac{1}{4}}\sqrt{|a|}}{\rm erf}\left(\frac{\alpha^{\frac{1}{4}}\sqrt{b}}{D^{\frac{1}{4}}}\right)e^{b\sqrt{\frac{\alpha}{D}}} \\
       &-\frac{\sqrt{2 b}}{\sqrt{|a|}}+\frac{D^{\frac{1}{4}}\sqrt{\pi}}{2 \sqrt{2}\alpha^{\frac{1}{4}}\sqrt{|a|}}{\rm erfi}\left(\frac{\alpha^{\frac{1}{4}}\sqrt{b}}{D^{\frac{1}{4}}}\right)e^{-b\sqrt{\frac{\alpha}{D}}},
  \end{split}
\end{equation*}
where ${\rm erfi}(x)$ is the imaginary error function. Finally, for the return under the action of the harmonic force, we have
\begin{equation*}
	 \langle t_f \rangle=\left(\frac{1}{\alpha}+\frac{\pi}{2\sqrt{k}}+\frac{1}{\lambda}\right)\Big(e^{b\sqrt{\frac{\alpha}{D}}}-1\Big).
\end{equation*}
The results of $\langle t_f \rangle$ are also verified in Fig. \ref{MFPT}, and a comparison with the MFPT of the process with instantaneous return considered in \cite{Puigdel} is also made in Fig. \ref{MFPT}.

%
%
%


\section{Conclusion}\label{sec 4}


We consider the process with three phases, which are movement phase, return phase, and rest phase. The three different methods of returning considered in this paper indicate the non-instantaneous reset. First for the general case, the PDF of overall process and PDF of return time  have been obtained. Then we further derive the asymptotic forms of MSD and $n$th moment of return time for each way of returning, i.e., return at a constant velocity or acceleration, or under the influence of harmonic force.



When the propagator function is Gaussian distribution and the particle returns at a constant velocity $v$, roughly speaking, there is always a competition between movement phase and rest phase; if the movement time follows exponential distribution, and the rest time has finite average, then the $\langle x^2(t) \rangle$ is a constant and decreases in $v$ for sufficiently long time, while for the case of divergent average of rest time, such as power-law distributed rest time, $\langle x^2(t) \rangle$ decreases in time $t$ and $v$. Similar phenomena can also be observed when the particle returns at constant acceleration $|a|$, and $\langle x^2(t) \rangle$ decreases in $|a|$. For the case of harmonic force with parameter $k$, when both of movement time and rest time are exponentially distributed with averages $1/\alpha$ and $1/\lambda$, then the overall MSD $\langle x^2(t)\rangle$ is closely related to $\alpha$, $\lambda$ and $k$, specifically, $\langle x^2(t) \rangle$ is decreasing in $k$ when $\alpha>\lambda$ otherwise it is increasing in $k$. Whereas, if the rest time has divergent average, then $\langle x^2(t) \rangle$ is simply decreasing in $t$ and $k$. For the case of power-law distributed movement time with power $-1-\alpha$, when the rest time has average  or power-law distributed \eqref{pw-law_PDF} with $\alpha<\beta$, then $\langle x^2(t) \rangle \sim t$, while for the case $\alpha > \beta$, it can be obtained that $\langle x^2(t) \rangle \sim t^{1-\alpha +\beta}$ indicating subdiffusion. Besides the PDF and $n$th moment of return time for each case are also calculated. Next we consider that the particle in movement phase follows dynamic of L\'evy walk, where the L\'evy walk behaves like ballistic diffusion and its walking duration follows power-law distribution with power $-\gamma-1$ and $\gamma=1/2$. If both of movement time and rest time have finite averages, then the overall MSD approximately behaves like a constant for each different kind of returns, while if the rest time follows power-law distribution in \eqref{pw-law_PDF}, then $\langle x^2(t)\rangle \sim t^{\beta-1}$ with $0<\beta<1$, indicating the decreasing of $\langle x^2(t)\rangle$ in $t$. If the movement time is power-law distribution with power $-1-\alpha$, then the ballistic diffusion begins to dominate the overall process. Specifically, the ballistic behavior, i.e., $\langle x^2(t)\rangle \sim t^2$, emerges when the rest time is exponentially distributed or power-law distributed like \eqref{pw-law_PDF} with $\alpha<\beta$, and for $\alpha>\beta$ there exists $\langle x^2(t)\rangle \sim t^{2+\beta-\alpha}$ indicating super-diffusion.

%

The MFPT of a given semi-infinite interval has also been considered in this paper for Brownian motion with exponentially distributed movement time and rest time. It is found that it has a close relation with the amount of return velocity, or return acceleration, or with the parameter of the  harmonic force. The smaller of return velocity, return acceleration, or the harmonic parameter is, the longer time is needed to arrive at the boundary of the semi-infinite interval, which is consistent with the intuition.

\section*{Acknowledgements}

This work was supported by the National Natural Science Foundation of China under Grant No. 12071195, and the AI and Big Data Funds under Grant No. 2019620005000775.

\begin{appendix}

\section{A brief introduction of the confluent hypergeometric function of the second kind}\label{Appen A}

The confluent hypergeometric function of the second kind $U(a,b;z)$ was introduced by Tricomi, also called Tricomi confluent hypergeometric function. Here we take $a$ and $b$ to be real numbers, and the $U(a,b;z)$ is defined by
\begin{equation}\label{a1}
\begin{split}
   U(a,b;z) =&\frac{\Gamma(1-b)}{\Gamma(a+1-b)}M(a,b;z)\\
     & + \frac{\Gamma(b-1)}{\Gamma(a)}z^{1-b}M(a+1-b,2-b;z),
\end{split}
\end{equation}
where $M(a,b;z)$ is the confluent hypergeometric function of the first kind.

As $|z|\rightarrow 0$, $M(a,b;z)\rightarrow 1$, we have when $|z|\rightarrow 0$,
\begin{equation}\label{a2}
   U(a,b,z) \sim \frac{\Gamma(1-b)}{\Gamma(a+1-b)}+
    \frac{\Gamma(b-1)}{\Gamma(a)}z^{1-b}.
\end{equation}

Moreover, the confluent hypergeometric function of the second kind $U(a,b;z)$ can be represented as an integral for $a>0$,
\begin{equation}\label{a3}
  U(a,b;z)=\frac{1}{\Gamma(a)}\int_{0}^{\infty}e^{-z t}t^{a-1}(1+t)^{b-a-1}dt.
\end{equation}
The Kummer transformation states that
\begin{equation}\label{a4}
  U(a,b;z)=z^{1-b}U(1+a-b,2-b;z).
\end{equation}

\section{Derivations of (\ref{surhatGb})}\label{app_b}

We take Laplace transform on \eqref{4.2}, and solve the equation of $\widehat{G}_b(s)$ in Laplace space. Obviously, the Laplace transform of the first term on r.h.s. of \eqref{4.2} is $\mathscr{L}_{t \to s}\left\{\Psi_m(t) H_b(t)\right\}$, and the Laplace transform of the second term is
\begin{widetext}

\begin{equation*}
	\begin{split}
		&\mathscr{L}_{t\to s}\left\{\int_{0}^{t} \psi_m (u) du \int_{-\infty}^{b} h_b(x,u) dx \int_{t-u}^{\infty} \delta\left(\theta - R^{-1}(0;x)\right)d\theta \right\}\\
		& = \int_{0}^{\infty} e^{-s t} dt \int_0^t \psi_m (u) du \int_{-\infty}^{b} h_b(x,u) dx \int_{t-u}^{\infty} \delta\left(\theta - R^{-1}(0;x)\right)d\theta \\
		& = \int_{0}^{\infty} \psi_m(u) du \int_{-\infty}^b h_b(x,u) dx \int_u^{\infty} e^{-st} dt\int_{t-u}^{\infty} \delta\left(\theta-R^{-1}(0;x)\right)d\theta\\
		&= \int_{0}^{\infty} \psi_m(u) du \int_{-\infty}^b h_b(x,u) dx \int_{0}^{\infty} d\theta \int_{u}^{u+\theta} e^{-st}  \delta\left(\theta-R^{-1}(0;x)\right) dt \\
		& \sim \int_{0}^{\infty} \psi_m(u) du \int_{-\infty}^b h_b(x,u) dx \int_{0}^{\infty}  \theta  \delta\left(\theta-R^{-1}(0;x)\right) d\theta\\
		& = \int_0^\infty \theta \varphi_b(\theta)d\theta = -\hat{\varphi}_b'(s)|_{s=0},
	\end{split}
\end{equation*}
where the asymptotic form is taken when $s$ is small enough. The Laplace transform of the third term on the r.h.s. of \eqref{4.2} is
\begin{equation*}
	\begin{split}
		&\mathscr{L}_{t\to s}\left\{\int_{0}^{t} \psi_m (u) du \int_{-\infty}^{b} h_b(x,u) dx \int_0^{t-u} \delta\left(\theta - R^{-1}(0;x)\right) \Phi_r(t-u-\theta)  d\theta \right\}\\
		&=\int_0^\infty e^{-s t} dt \int_{0}^{t} \psi_m (u) du \int_{-\infty}^{b} h_b(x,u) dx \int_0^{t-u} \delta\left(\theta - R^{-1}(0;x)\right) \Phi_r(t-u-\theta)  d\theta \\
		&= \int_0^\infty \psi_m (u) du \int_{-\infty}^{b} h_b(x,u) dx \int_0^\infty e^{-s (\theta + u)} d\theta \int_0^\theta \delta\left(\eta - R^{-1}(0;x)\right) \Phi_r(\theta-\eta)  d\eta \\
		&=\int_{-\infty}^b dx \mathscr{L}_{u \to s} \left\{\psi_m(u) h_b(x,u) \right\} e^{-s R^{-1}(0;x)} \widehat{\Phi}_r(s)\\
		&\sim \mathscr{L}_{u \to s} \left\{\psi_m(u) H_b(u) \right\} \widehat{\Phi}_r(s),
	\end{split}
\end{equation*}
where the asymptotic form is still obtained for small enough $s$. Similarly, for the last term on r.h.s. of \eqref{4.2}, the asymptotic behavior in Laplace space for small $s$ is
\begin{equation*}
	\begin{split}
		& \mathscr{L}_{t\to s}\left\{\int_{0}^{t} \psi_m(u) du \int_{-\infty}^{b} h_b(x,u) dx\int_{0}^{t - u} \delta\left(\theta - R^{-1}(0;x) \right) d\theta \int_{0}^{t-u-\theta} \phi_r(\eta) G_b(t - u - \theta - \eta) d\eta \right\} \\
		& = \int_{-\infty}^b dx \mathscr{L}_{t \to s}\left\{\psi_m(t) h_b(x,t) \right\} e^{-s R^{-1}(0;x)} \hat{\phi}_r (s) \widehat{G}_b(s)\\
		& \sim \mathscr{L}_{t \to s}\left\{\psi_m(t) H_b(t) \right\}  \hat{\phi}_r (s) \widehat{G}_b(s)
	\end{split}
\end{equation*}
Then from \eqref{4.2} we have the asymptotic behavior of $\widehat{G}_b(s)$ for small $s$, i.e.,
\begin{equation*}
   \widehat{G}_b(s)  \sim \frac{\mathscr{L}_{t\rightarrow s}\left\{\Psi_m(t) H_b(t)\right\}-\hat{\varphi}_b'(s)|_{s=0}}{1-\hat{\phi}_r(s)\mathscr{L}_{t\rightarrow s}\{\psi_m(t)H_b(t)\}}  +\frac{\widehat{\Phi}_r(s)\mathscr{L}_{t\rightarrow s}\{\psi_m(t)H_b(t)\}}{1-\hat{\phi}_r(s)\mathscr{L}_{t\rightarrow s}\{\psi_m(t) H_b(t)\}}.
\end{equation*}
\end{widetext}

\end{appendix}

\bibliography{ref}

\end{document}